\documentclass[a4paper,reqno]{amsart}

\usepackage[text={5.0in,8.0in},centering,top=30mm]{geometry}

\usepackage[usenames,dvipsnames,svgnames,table]{xcolor}
\definecolor{citegreen}{rgb}{0.00,0.70,0.30}
\usepackage[colorlinks=true,
            linkcolor=blue,
            urlcolor=blue,
            citecolor=blue]{hyperref}
\usepackage{amsrefs}
\usepackage{amssymb}

\usepackage{mathrsfs}
\usepackage{mathtools}

\usepackage{stmaryrd}

\usepackage{xspace}

\usepackage{pgf}
\usepackage{tikz}

\usetikzlibrary{arrows,automata}

\usetikzlibrary{calc}

\DeclareMathAlphabet{\mathpzc}{OT1}{pzc}{m}{it}

\usepackage{epsfig}
\usepackage{enumerate}
\usepackage{graphicx}

\numberwithin{equation}{section}
\theoremstyle{plain}
\newtheorem{theorem}{Theorem}
\newtheorem{prop}{Proposition}[section]

\newtheorem{lemma}{Lemma}[section]

\newtheorem{corollary}{Corollary}[section]

\theoremstyle{remark}
\newtheorem{remark}{Remark}

\newtheorem*{quest*}{Question}
\newtheorem*{remark*}{Remark}
\theoremstyle{remark}

\theoremstyle{definition}
\newtheorem{definition}{Definition}[section]
\newtheorem*{definition*}{Definition}
\newtheorem*{notation*}{Notation}
\newtheorem*{notations*}{Notations}

\DeclareMathOperator{\Tr}{Tr\,}

\providecommand{\D}{\mathbb}

\providecommand{\R}{\mathrm}

\newcommand{\eu}{\mathrm{e}}
\newcommand{\ii}{\mathrm{i}}

\def\ball{\mathrm{B}}

\def\ballj{\mathrm{B}_{L_j}}
\def\balljone{\mathrm{B}_{L_{j+1}}}

\DeclareMathOperator{\dist}{dist}

\DeclareMathOperator{\card}{card}

\DeclareMathOperator*{\essup}{ess\,sup}
\DeclareMathOperator*{\supp}{supp}

\DeclareMathOperator{\Unif}{Unif}

\DeclareMathOperator{\one}{\mathbf{1}}

\DeclareMathOperator{\sign}{sign}

\DeclareMathOperator{\diam}{{\rm diam}}

\def\mytimes{\operatornamewithlimits{\hbox{\huge$\times$}}}

\def\cdott{\cdot\,}

\def\sgomth#1#2{{\Sigma(H_{\ball_{#1}(#2)}(\om,\th))}}

\def\lr#1{{\langle #1\rangle}}

\def\DIV{{\rm\textbf{(DIV)}}\xspace}
\def\LVB{{\rm\textbf{(LVB)}}\xspace}

\def\UPA{{\rm\textbf{(UPA)}}\xspace}

\def\sparsez{{\rm \textbf{Sparse($\mathbf{0}$)}}\xspace}
\def\sparseL#1{{\rm \textbf{Sparse($\mathbf{#1}$)}}\xspace}
\def\sparsepar#1{{\rm \textbf{Sparse($\mathbf{L_{#1}}$)}}\xspace}
\def\sparsej{{\rm \textbf{Sparse($\mathbf{L_{j}}$)}}\xspace}
\def\sparsejone{{\rm \textbf{Sparse($\mathbf{L_{j+1}}$)}}\xspace}

\def\EmNS{$(E,m)${\rm-NS}\xspace}
\def\EmS{$(E,m)${\rm-S}\xspace}

\def\icirc{{i_\circ}}

\def\rc{\mathrm{c}}

\def\lam{{\lambda}}

\def\th{\vartheta}
\def\Th{{\rm \Theta}}
\def\eps{\epsilon}
\def\ffi{\varphi}

\def\taujl{\tau_{j,l}}
\def\tauminusl{\tau_{-1,l}}

\def\ER{$E$-R\xspace}

\def\mbad{$m${\rm-bad}\xspace}
\def\mgood{$m${\rm-good}\xspace}

\def\ER{$E${\rm-R}\xspace}
\def\ENR{$E${\rm-NR}\xspace}

\def\tb{{ \tilde{b} }}

\def\hk{{\widehat{k}}}

\def\hth{{{\widehat{\th}}}}

\def\BX{\mathbf{X}}
\def\BY{\mathbf{Y}}
\def\BZ{\mathbf{Z}}

\def\Const{{\rm{Const}}}

\def\tnL{{\widetilde{n}(L)}}
\def\tn{{\widetilde{n}}}

\def\tA{{\widetilde{A}}}
\def\tNL{{\widetilde{N}(L)}}
\def\tN{{\widetilde{N}}}

\def\Nj{{\widetilde{N}_j}}
\def\Nminus{{\widetilde{N}_{-1}}}

\def\DC{\D{C}}

\def\DP{\D{P}}
\def\DQ{\D{Q}}
\def\DR{\D{R}}
\def\DT{\D{T}}
\def\DZ{\D{Z}}
\def\DN{\D{N}}

\def\cB{\mathcal{B}}

\def\csC{\mathscr{C}}

\def\cC{\mathcal{C}}

\def\cF{\mathcal{F}}

\def\cH{\mathcal{H}}
\def\cI{\mathcal{I}}
\def\cK{\mathcal{K}}

\def\cL{\mathcal{L}}
\def\cM{\mathcal{M}}

\def\cP{\mathcal{P}}

\def\cT{\mathcal{T}}
\def\cV{\mathcal{V}}

\def\CZ{\mathcal{Z}}

\def\rA{{\R{A}}}

\def\rM{{\R{M}}}

\def\rP{{\R{P}}}

\def\hx{\hat{x}}
\def\hy{\hat{y}}
\def\hX{\hat{X}}

\def\hlam{\hat{\lambda}}

\def\be{\begin{equation}}
\def\ee{\end{equation}}
\def\ba{\begin{array}{l}}
\def\ea{\end{array}}
\def\bal{\begin{aligned}}
\def\eal{\end{aligned}}

\def\ble{\begin{lemma}}
\def\ele{\end{lemma}}

\def\bco{\begin{cor}}
\def\eco{\end{cor}}

\def\bpr{\begin{prop}}
\def\epr{\end{prop}}

\def\bre{\begin{remark}}
\def\ere{\end{remark}}

\def\btm{\begin{theorem}}
\def\etm{\end{theorem}}

\def\bde{\begin{definition}}
\def\ede{\end{definition}}

\def\ffi{{\varphi}}

\def\fB{\mathfrak{B}}
\def\fF{\mathfrak{F}}

\def\fh{\mathfrak{h}}
\def\fs{\mathfrak{s}}

\def\om{{\omega}}
\def\Om{{\Omega}}

\def\eps{\epsilon}

\def\Lam{{\Lambda}}
\def\lam{{\lambda}}

\def\thnk{{\th_{n,k}}}

\def\ffink{{\ffi_{n,k}}}

\def\fsnk{{\fs_{n,k}}}

\def\Cnk{{C_{n,k}}}
\def\Cpnk{{C^{+}_{n,k}}}
\def\Cmnk{{C^{-}_{n,k}}}

\def\Thinf{{\Th^{(\infty)}}}
\def\ThinfM{{\Th^{(\infty)}_{\rm M}}}

\def\Thminus{{\Th^{(-1)}}}

\def\Sep#1{{{\rm\mathbf{Sep}}\left[#1\right]}}
\def\Sepb#1{{{\rm\mathbf{Sep}}\big[#1\big]}}
\def\SepB#1{{{\rm\mathbf{Sep}}\Big[#1\Big]}}

\def\promth#1{{\DP^{\Om\times\Th}\left\{\,#1\,\right\}}}

\def\cond{\,\big|\,}
\def\Bcond{\,\Big|\,}
\def\prth#1{{\DP^{\Th}\left\{\,#1\,\right\}}}
\def\prthp{{\DP^{\Th}}}

\def\prTh#1{{\DP^{\Th}\left\{\,#1\,\right\}}}

\def\esmth#1{\D{E}^{\Theta}\left[\, #1\, \right]}
\def\esmomth#1{\D{E}^{\Omega\times\Theta}\left[\, #1\, \right]}

\def\pt{\partial}
\def\half{\frac{1}{2}}

\def\truc#1#2#3{\smash{\mathop{\,\, #1 \,\, }\limits^{#2}_{#3}}}

\def\tto#1{\smash{\mathop{\,\,\,\, \longrightarrow \,\,\,\, }\limits_{#1}}}

\def\myset#1{{\left\{\,#1\,\right\}}}

\def\mymax#1{{ \truc{\max} {} {#1}}}

\def\be{\begin{equation}}
\def\ee{\end{equation}}
\def\ba{\begin{array}{l}}
\def\ea{\end{array}}
\def\bal{\begin{aligned}}
\def\eal{\end{aligned}}

\def\ble{\begin{lemma}}
\def\ele{\end{lemma}}

\def\bre{\begin{remark}}
\def\ere{\end{remark}}

\def\btm{\begin{theorem}}
\def\etm{\end{theorem}}

\def\bde{\begin{definition}}
\def\ede{\end{definition}}

\def\bpr{\begin{prop}}
\def\epr{\end{prop}}

\def\bco{\begin{corollary}}
\def\eco{\end{corollary}}

\definecolor{redd}{rgb}{0.95,0.2,0.2}
\definecolor{gris}{rgb}{0.9,0.9,0.9}

\definecolor{cmgray}{rgb}{0.7,0.7,0.7}

\definecolor{cmblue}{rgb}{0.2,0.5,0.8}

\begin{document}

\title[Uniform localization and simple spectra for ``haarsh'' potentials]
{Uniform Anderson localization,\\ unimodal eigenstates and simple spectra
\\in a class of ``haarsh'' deterministic potentials}

\author[V. Chulaevsky]{Victor Chulaevsky}


\address{D\'{e}partement de Math\'{e}matiques\\
Universit\'{e} de Reims, Moulin de la Housse, B.P. 1039\\
51687 Reims Cedex 2, France\\
E-mail: victor.tchoulaevski@univ-reims.fr}

\date{\today}

\begin{abstract}
We study a particular class of  families of multi-dimensional lattice Schr\"{o}\-dinger operators
with deterministic (including quasi-periodic) potentials generated by the "hull" given by an orthogonal
series over the Haar wavelet basis on the torus, of arbitrary dimension, with expansion coefficients
considered as independent parameters.
In the strong disorder regime, we prove Anderson localization for generic operator families,
using a variant of the Multi-Scale Analysis, and show that all localized eigenfunctions are
\emph{unimodal} and feature
\emph{uniform} exponential decay  away from their respective localization centers.
Using the Klein--Molchanov argument and a variant of the Minami estimate for deterministic potentials,
we  prove the simplicity of the spectrum in our model.

\vskip5mm
\textcolor{blue}
{\textbf{NOTE}: This text completes our earlier manuscript (\texttt{math-ph/0907.1494}),
originally uploaded in 2009 and revised in 2011,
which is kept in \textbf{arXiv} in a reduced form, merely to avoid broken references in earlier works.
Compared to [\texttt{math-ph/0907.1494}], we add the results on unimodality of the eigenstates,
uniform dynamical localization, and simplicity of p.p. spectra.
\vskip3mm
Compared to  earlier versions of this preprint, the presentation has been
adapted to the future extension of the main results (uniform localization, unimodality
of the eigenfunctions) to the multi-particle Anderson Hamiltonians
with a nontrivial interaction between the particles, which we plan to publish in a
forthcoming paper.
%
}

\end{abstract}

\maketitle


\section{Introduction. The model and the main results.}\label{intro}

We study spectral properties of finite-difference operators, usually called
discrete (or lattice) Schr\"{o}dinger operators (DSO), of the form
\be\label{eq:DSO.intro}
(H(\om;\th) f)(x) = \sum_{y:\, \|y-x\|=1} f(y) + gV(x;\om;\th) f(x),\; x,y\in \DZ^d, \, g\in\DR,
\ee
where $\om$ and $\th$ are parameters, the role of which we explain below.

%
In mathematical modeling of disordered quantum systems, it makes more sense to study
not an individual operator, but an entire family
$H(\omega)$ labeled by the points of the phase space of a dynamical system on
some probability space. Moreover, it is often convenient
(but not always necessary)
to assume ergodicity of the dynamical
system in question. The usual approach to the notion of ergodic ensemble of
operators in $\ell^2(\DZ^d)$ is as follows: one considers
an ergodic dynamical system $T$ with discrete time $\DZ^d$, $d\geq 1$, on
a probability space $(\Omega, \cF, \DP)$,
and a measurable mapping $H$ of the space $\Omega$ into the space of
operators (for example, bounded) acting in the Hilbert space ${\cH} = l^2(\DZ^d)$ and satisfying
for every
$x\in\DZ^d$:
$$
H( T^x(\omega) ) = U^{-x} H(\omega) U^x,
$$
where $(U^x f)(y) = f(y-x)$ are the conventional, unitary shift operators.
In particular, the DSO \eqref{eq:DSO.intro}
is obtained by setting
$
H(\om) = \Delta + V(x;\om)
$,
where $(\Delta f)(x) = \sum_{|y-x|=1} f(y)$,  and $V(\cdot\,;\om)$ is the operator
of multiplication by the function
\be\label{eq:V.v.T}
x \mapsto V(x;\om) = v(T^x \om),
\ee
where the function $v:\,\Om\to\DR$ will be called the \textit{hull} of the potential $V$.

A rich and  interesting class of \emph{quasi-periodic} potentials, e.g., in one dimension,
is obtained when $\Om$ is the torus $\DT^1$
endowed with the  Haar measure $\DP$, and  the  dynamical system on $\Om$ is  given by
$
T^x:\, \om  \mapsto  \om + x \alpha, \;\; \om\in \DT^1,
$
and $\alpha\in\DR\setminus\DQ$.
This dynamical system is well-known to be ergodic. Taking a function $v:\DT^1 \to \DR$, we can define an ergodic family of quasi-periodic potentials $V:\DZ\to \DR$ by $V(x;\om) := v(T^x \om)$. Multi-dimensional quasi-periodic potentials on $\DZ^d$ can be constructed in a similar way (with the help of $d$ incommensurate frequency vectors $\alpha^{j}\in\DR^\nu, j=1, \ldots, d$).
In the  case where  $v(\om) = g\cos (2\pi \om)$, $g\in\DR$,  $\alpha\in\DR\setminus \DQ$,
the DSO $H(\om)$ with the potential of the form \eqref{eq:V.v.T} is called  Almost Mathieu or Harper's operator.

Sinai \cite{Sin87}  and Fr\"{o}hlich et al. \cite{FSW90} proved Anderson localization
for a class of the DSO with the ``cosine-like'' potential; more precisely, the hull $v:\DT^1\to\DR$
was assumed to be of the class $\cC^2(\DT^1)$ and have exactly two extrema, both non-degenerate.
Operators with several basic frequencies (i.e., $\om\in\DT^\nu$, $\nu >1$) were studied
in \cite{CSin89} ($\nu=2$),
and later in a cycle of papers by Bourgain, Goldstein and Schlag, for various dynamical systems
on a torus $\Om=\DT^\nu$, $\nu\le 2$,
where the hull $v(\om)$ was assumed analytic; see, e.g., \cite{BG00}, \cite{BGS01}, \cite{BS00}.
More recently, Chan \cite{Chan07} used a parameter exclusion technique (different from ours) to establish the  localization for quasi-periodic operators with sufficiently non-degenerate hull $v\in \csC^3(\DT^1)$.

Note that the number of rigorous results on Anderson localization for almost-periodic
and, more generally, deterministic families of random operators remains rather limited,
particularly in dimension $d>1$,
compared to the wealth of results for Schr\"{o}dinger-type operators with
IID or weakly correlated random potentials.

Among recent results most closely related to the topic of the present paper, we refer
to the works by Damanik and Gan \cite{DG10,DG12} who proved uniform localization for a
class of one-dimensional operators with limit-periodic potential.

\vskip1mm

$\blacklozenge$ In the present paper, we consider \emph{parametric families} of hulls on the phase space
$\Om$, $\{v(\cdott;\th), \th\in\Th\}$, labeled by
elements $\th$ of an auxiliary set $\Th$ which we endow with the structure of a probability space;
the construction is described in Sect.~\ref{ssec:randelettes} and
\ref{sec:partitions}. It is this specific construction which allows us to
prove our main result on genuinely uniform Anderson localization for typical values
of $\th\in\Th$ (see Theorem \ref{thm:Main} in Sect.~\ref{ssec:Main.res}).
We encapsulate the main requirement for the underlying dynamical system,
generating the deterministic random potential, in one mild condition -- "Uniform
Power-law Aperiodicity" (\UPA; cf. \eqref{eq:cond.UPA} in Sect.~\ref{ssec:UPA}).

$\blacklozenge$ It is to be emphasized that the ergodicity of the dynamical system
is not required \emph{per se }for our proof of localization. However, in the case where
$T$ is generated by the shifts of the torus $\DT^1$,  aperiodicity implies
topological transitivity, hence  ergodicity of $T$. In fact, for the toral shifts,
the condition \UPA reads as the Diophantine condition on the frequency vectors.

\vskip1mm

Our class of models features
unusually strong localization properties, similar to those of the celebrated Maryland model,
discovered and studied by physicists Fishman et al. \cite{FGP84}.
The potential in the Maryland model is  quasi-periodic and generated by the analytic hull
$$
\om \mapsto g \, \tan \pi\om, \;\; \om\in\DT^1 \cong [0,1) \subset \DR \hookrightarrow \DC,
$$
which admits a meromorphic continuation to the complex plane. Its restriction to $\DR$
is strictly monotone on the period (between two consecutive poles),
and this ultimately results in complete absence of ``resonances''
between distant sites on the lattice $\DZ^d$. In turn, this gives rise
to the exponentially localized eigenstates which are unimodal, i.e., cannot have multiple "peaks".

The notion of a "peak" actually makes sense
for the disorder amplitude $|g|\gg 1$: in this case, the Maryland operator has an
orthonormal eigenbasis
of exponentially fast decaying eigenfunctions $\psi_x$, labeled in a non-ambiguous and natural way by
the points $x\in\DZ^d$ so that
$$
\begin{aligned}
\inf_{x\in \DZ^d} \; |\psi_x(x)|^2 &\ge 1 - f(|g|) > {\textstyle \half },
\;\; f(|g|) \tto{|g|\to\infty} 0.
\end{aligned}
$$
In other words, for $|g|\gg 1$,
the eigenbasis for $H(\om)$ is a small-norm perturbation of the
standard delta-basis in $\ell^2(\DZ^d)$; this would be, of course,  an event of probability 0 for
the random Anderson Hamiltonians.

Another particularity of the Maryland model, rigorously proven in independent
mathematical works by Figotin and Pastur \cite{FiP84} and by Simon \cite{Sim85},
is the \emph{non-perturbative} complete exponential localization: it occurs for
any, arbitrarily small amplitude of disorder $|g|>0$. With the exception for this particular feature,
the "unimodal", uniform exponential localization was extended by Bellissard et al.
\cite{BLS83}
to the class of meromorphic hulls with a real period, strictly monotone on the period.
The proof in \cite{BLS83} is a linear version of the KAM
(Kolmogorov--Arnold--Moser)
method, which requires the parameter
$|g|^{-1}$ to be small enough for the inductive procedure to succeed, so it remains yet unknown
if the complete localization occurs in the BLS-class for arbitrarily weak disorder.

The class of deterministic Anderson
models considered in this paper features the same complete unimodality
of the eigenbasis, i.e.,  genuinely uniform decay of all eigenfunctions, and not just semi-uniform
(often referenced to as the SULE property: Semi-Uniformly Localized Eigenfunctions).
This class also has important particularities:

\begin{enumerate}
  \item The class of the underlying dynamical systems, representing the disorder from the traditional point of view, is not limited to quasi-periodic or, more generally, almost-periodic systems. This is explained by the fact that the "dynamical disorder" plays here a subordinate, indeed minor role
      in the localization, while the dominant role is given to the "parametric disorder", responsible for the decay of eigenfunctions.

  \item The uniform decay of eigenfunctions occurs for \emph{all} phase points of
      the dynamical system, and not just Lebesgue-almost all, as in many quasi-periodic systems, e.g.,
      for the Almost Mathieu operators. On the other hand, it occurs only for a subset of the parameter
      set, labeling the hulls. The measure of the excluded subset decays as $|g|\to\infty$.
      In other words, we prove locali\-zation for a.e. $\th\in\Th$ and all $\om\in\Om$,
      but with $|g|\ge g^*(\th)$.

  \item The hulls under consideration are, speaking pictorially, ``made out of flat pieces"
  (viz. composed of Haar wavelets), while in most models, one usually
  had to make special efforts to avoid ``flat''
  components of the random or deterministic hulls. Albeit the hulls ultimately become non-flat, they
  are piecewise-constant at every step of the inductive approximation
  procedure, and \emph{this} is precisely what
  gives rise to the \emph{uniform} exponential localization.
\end{enumerate}

\vskip1mm
In this work, as in \cite{C11c},
we often use the term \emph{random}, sometimes putting it in quotes, and this might create the
illusion that the operators with deterministic -- e.g., quasi-periodic -- potentials, considered here,
are somehow perturbed by a masterly hidden random noise.
We do not add, or otherwise introduce, any IID or weakly correlated noise
in the potential, which always remains deterministic, with stochastic
properties\footnote{As the matter of fact, we do not make use of any stochastic properties
of the underlying dynamics, other than the "Uniform Power-law Aperiodicity" (cf. \eqref{eq:cond.UPA}).}
induced exclusively by the underlying dynamical system.
For example, if $\{T^x, x\in\DZ^d\}$ is generated by incommensurate shifts of the torus,
the obtained potentials are always quasi-periodic, thus feature the weakest possible ergodic properties.
Yet, it is true that many techniques used in the proof of localization come from the
conventional theory of random Anderson Hamiltonians.

One particularly important advantage of the probabilistic language and tools is that we can prove
Minami-type estimates, of all orders,
for generic deterministic operator ensembles. Combined with the
Klein--Molchanov argument (cf. \cite{KM06}), this results in the proof of simplicity of the pure point
spectrum, for every (and not just a.e.) phase point of the underlying dynamical system. To the best of the author's knowledge, this is the first result of such
kind for a  large class of deterministic operators. It is not related to the unimodality
of the eigenstates; in a forthcoming work, following the path laid down in \cite{C11c},
we will extend it to a more general class
of deterministic DSO with hulls of any finite smoothness, where the respective Hamiltonians
feature only the SULE property, and the eigenstates are not unimodal.
\vskip1mm

Our main results are presented in Sect.~\ref{ssec:Main.res}.
\vskip1mm

Technically speaking, the most tedious analysis is required to establish analogs of the Wegner estimate,
and infer from them the unusual -- uniform -- lower bounds
on the ``small denominators'', or ``resonances''.
Once such bounds are obtained, the derivation of the Anderson localization
becomes quite simple and ``soft'' (cf. Sect.~\ref{sec:MSA}); the reader will see that
it is actually simpler than for the Anderson Hamiltonians with IID random potential.

\subsection{Requirements for the dynamical system}\label{ssec:UPA}

For the sake of clarity,
we consider in this paper only the case where $\Om = \DT^\nu$, $\nu\ge 1$, and it is convenient
to define the distance $\dist_\Om[\om', \om'']$ as follows: for
$\om'=(\om'_1, \ldots, \om'_\nu)$ and $\om''=(\om''_1, \ldots, \om''_\nu)$,
$$
\dist_\Om[\om', \om'']
:= \max_{1 \le i \le \nu} \dist_{\DT^1}[\om'_i, \om''_i],
$$
where $\dist_{\DT^1}[\,\cdot\,,\,\cdot\,]$
is the conventional distance on $\DT^1 = \DR^1/\DZ^1$.
With this definition, the diameter of a cube of side length $r$ in $\DT^\nu$ equals $r$, for any dimension
$\nu\ge 1$. The main reason for the choice of the phase space $\Om=\DT^\nu$ is that the parametric families of ensembles of potentials $V(x;\om;\th)$ are fairly explicit in this case, and this allows one
to construct  quasi-periodic operators.

We assume that the underlying dynamical system $T$ (generating the potential)
satisfies the condition of Uniform Power-law Aperiodicity:

\UPA \quad $\exists\, A, C_A\in\DN^* \; \;\forall\, \om\in\Om \;
  \forall\, x, y\in\DZ^\nu \text{ such that } x\ne y $
\be\label{eq:cond.UPA}
\begin{array}{lc}
\quad \dist_{\Om}(T^x \om, T^y \om) \ge C^{-1}_A |x - y |^{-A},
\end{array}
\ee
and the condition of tempered local divergence of trajectories:
\par

\DIV \quad
$\exists\, A', C_{A'}\in\DN^*\; \;\forall\, \om, \om'\in\Om \;\forall\, x\in\DZ^\nu\setminus\{0\}$
\be\label{eq:cond.DIV}
\begin{array}{lc}
\quad \dist_{\Om}(T^x \om, T^x \om') \le C_{A'}\, |x|^{A'} \dist_{\Om}(\om, \om').
\end{array}
\ee

\bre
It is not difficult to see that both \UPA and \DIV rule out strongly mixing dynamical systems like
the hyperbolic toral automorphisms (while the skew shifts of tori are still allowed).
This certainly looks quite surprising, but it has to be emphasized
that our proof is oriented towards the dynamical systems with the \emph{weakest} stochasticity. In a manner
of speaking, we actually need that the dynamical system ``do not interfere'' with the ``randomness''
provided by the parametric freedom in the choice of the sample potential $V(\cdot;\om;\th)$. As to
the mixing systems, their intrinsic randomness is to be used in the proof of localization
in a different way; this puts them beyond the scope of the present paper. Note, however, that
the localization properties of deterministic DSO with strongly mixing potential should, in our opinion,
be similar to those of the genuinely random DSO. In particular, we believe that
the uniform decay and unimodality of the eigenfunctions should not occur for the DSO with sufficiently
strongly mixing potentials.
\ere

For the rotations of $\DT^\nu$,
\DIV holds trivially, since $T^x$ are isometries,
and
\UPA reads as the Diophantine condition for the frequencies.

\subsection{The Local Variation Bound}

We often work with lattice cubes $\ball_L(u) := \{x\in\DZ^d:\, |x-u| \le L\}$, $L\ge 0$;
for $y=(y_1, \ldots, y_d)\in\DZ^d$, $|y|$ stands for the max-norm, $|y|:= \max_{i}|y_i|$.

Following \cite{C12a}, we introduce now a hypothesis on the random field
$v:\Om\times\Th\to\DR$ on $\Om$,
relative to the probability space  $(\Th,\fB,\prthp)$, which is logically independent of the
particular construction given in Sect.~\ref{ssec:randelettes}. Later we will show that it
holds true for the hulls constructed with the help of the randelette expansions in
Sect.~\ref{ssec:randelettes}.

\vskip2mm
\LVB:
\emph{
Let $v:\Om\times\Th\to\DR$ be a measurable function on the product probability space
$(\Om\times\Th, \fF\times\fB,\DP\times\prthp)$.
There exists a family of sub-sigma-algebras $\fB_L \subset \fB$, $L\in\DN^*$, such that, conditional
on $\fF\times\fB_L$ (hence, with $\om\in\Om$ fixed), for any cube $\ball_{L^4}(u)$, the values
$\{V(x;\om;\th):=v(T^x \om;\th), \, x\in \ball_{L^4}(u)\}$, are (conditionally) independent and admit individual (conditional) probability densities $\rho_{v,x}(\cdot\,|\fF\times\fB_L)$, satisfying}
\be\label{eq:LVB}
\essup
\| \rho_{v,x}(\cdot\,|\fF\times\fB_L) \|_\infty \le C'' L^{B \ln L}, \;\; C''\in(0,+\infty).
\ee
\vskip2mm

It is readily seen that for the scaled random variables $(\om;\th) \mapsto g V(x;\om;\th)$
the assumption \eqref{eq:LVB} implies
\be
\essup
\| \rho_{gv,x}(\cdot\,|\fF\times\fB_L) \|_\infty \le C'' g^{-1} L^{B \ln L}, \;\; C''\in(0,+\infty).
\ee

This property allows us to prove satisfactory analogs of the Wegner (cf. Sect.~\ref{ssec:Wegner})
and Minami (cf. Sect.~\ref{sec:Minami})
estimates in finite cubes
of any size $L$.

\subsection{Lattice cubes and local Hamiltonians}

Given a DSO $H = \Delta + gV$, where $V:\DZ^d\to\DR$ and $g > 0$,
and a proper subset $\Lam\subsetneq \DZ^d$, we consider the restriction
$H_\Lam$ of $H$ to $\Lam$ defined as follows:
$H_\Lam = \one_\Lam H \one_\Lam \upharpoonright \ell^2(\Lam)$; here the indicator
function $\one_\Lam$ is identified with the multiplication operator by this function,
and also with the natural orthogonal projection from $\ell^2(\DZ)$ onto $\ell^2(\Lam)$.
$H_\Lam$ is usually considered as the discrete analog of the Schr\"{o}dinger operator
with Dirichlet boundary conditions, acting on functions $\psi$ vanishing outside $\Lam$.

\subsection{Randelette expansions: An informal discussion}
\label{ssec:randelettes}

In Ref. \cite{C11c} we introduced \textit{parametric families} of ergodic ensembles of operators
$\{H(\om;\th), \om\in\Om\}$ depending upon a parameter $\th\in\Th$ in an auxiliary space $\Th$.
As shows \cite{C11c}, it is convenient to endow $\Th$ with the structure of a probability
space, $(\Th, \fB, \prthp)$, in such a way that $\th$ be, in fact, an \textit{infinite} family of
IID random variables on $\Th$, providing an infinite number of auxiliary independent
parameters allowing to vary the
hull $v(\om;\th)$ \emph{locally} in the phase space $\Om$. We called such parametric
families \emph{grand ensembles}.

The above description is, of course, too general. In the framework of the DSO,
we proposed in  \cite{C11c} a more specific construction where $H(\om;\th) = H_0 + V(\cdot;\om;\th)$,
with $V(x;\om;\th) = V(T^x\om;\th)$ and
\be\label{eq:randelettes.1}
v(\om; \th) = \sum_{n=0}^\infty a_n \sum_{k=1}^{K_n} \th_{n,k} \ffi_{n,k}(\om),
\ee
where $\{\thnk, n\ge 0, \; 1 \le k \le K_n \}$  are IID random variables on $\Th$, and
$\ffink := (\ffink), n\ge 0, \; 1 \le k \le K_n<\infty$, are some functions on the phase space $\Om$ of the
underlying dynamical system $T^x$. Series of the form \eqref{eq:randelettes.1}
were called in \cite{C11c} \textit{randelette}  expansions, referring to the "random" nature of the
expansion coefficients and to the shape of $\ffi_{n,k}$ reminding the wavelets ("\emph{ondelettes}",
in French).

Putting the amplitude of the function $\ffink$ essentially
in the "generation" coefficient $a_n$, it is natural to assume
that $|\ffink(\om)|$ are uniformly bounded in $(n,k,\om)$.
Further, in order to control the potential  $V(T^x \om;\th)$ at any lattice site $x\in\DZ^d$
or, equivalently, at every point $\om\in\Om$, it is natural to require that for every $n\ge 1$,  $\Om$
be covered by the union of the sets where at least one function $\ffink$ is nonzero (and, preferably, not
too small).

In the next subsection, we make a specific choice for $\{a_n\}$ and $\{\ffink\}$.

Notice that the dynamics $T^x$ leaves $\th$ invariant.

\subsection{Lacunary ``haarsh'' randelette expansions}
\label{ssec:lacunary.radelettes}

A very particular and  interesting case is where the randelettes are simply Haar wavelets
with coefficients considered, formally,
as independent random variables relative to an auxiliary probability space
$(\Th,\fB,\prthp)$.
For example, if $\Om = \DT^1 = \DR/\DZ$, for $n=0$ we set $K_0 = 1$, $\ffi_{0,1}(\om) = 1$,
and for
$n\ge 1$, $1 \le k \le K_n = 2^n$,
$$
\ffink(\om) = \one_{\Cpnk}(\om)-\one_{\Cmnk}(\om),
$$
where
\be\label{eq:C.plus.minus.n.k}
\Cpnk = \left[\frac{k-1}{2^{n}}, \frac{k-1}{2^{n}} + \frac{1}{2^{n+1}}\right), \quad
\Cmnk = \Cpnk + \frac{1}{2^{n+1}},
\ee
so
\be\label{eq:Cnk.Cpnk.Cmnk}
\supp \ffi_{n,k} = \Cnk := \Cpnk \cup \Cmnk.
\ee
On the torus $\DT^\nu$ with $\nu>1$, the functions $\ffink$ are tensor products of the one-dimensional
Haar's wavelets, and $\Cnk := \supp \ffink$ are cubes in $\DT^\nu$ of side length $2^{-n}$,
of the form
$$
\Cnk = \mytimes_{j=1}^\nu \left[ \frac{k_j}{2^n}, \frac{k_j+1}{2^n}  \right);
$$
they define a partition of $\DT^\nu$ which we denote by $\cC_n$.

Furthermore, each of these cubes is
partitioned into $2^\nu$ sub-cubes of side length $2^{-n-1}$,
$\{C_{n,k;i}, \, i=1, \ldots, 2^\nu\}$,
on which $\ffink$ takes a constant value $\pm 1$;
we denote this value by $\fsnk(\om) \in \{-1,+1\}$, so that
\be\label{eq:ffink.fsnk.Cnk}
\ffink(\om) = \fsnk(\om) \one_{\Cnk}(\om).
\ee
Clearly, the cubes $C_{n,k;i}$ are elements of the finer partition $\cC_{n+1}$. Indeed,
similar to \eqref{eq:C.plus.minus.n.k}, we have
\be\label{eq:Cbullnk}
C_{n,k;i} = \mytimes_{j=1}^\nu
\left[ \frac{k_j}{2^n} + \frac{l_{j;i}}{2^{n+1}}, \frac{k_j}{2^n} + \frac{l_{j;i}+1}{2^{n+1}} \right),
\;\; l_{j,i}\in \{0, 1\},
\ee
where the combinations of the shifts $l_{j;i}$ determine $\sign( \fsnk(\cdot))$.

Next, consider a  family of IID random variables $\thnk$ on an auxiliary probability space
$(\Th,\fB,\prthp)$, uniformly distributed in $[0,1]$.

Finally, let
\be\label{eq:def.an.b}
a_n = 2^{-2bn^2}, \; n\ge 1, \;\; b>0,
\ee
with $b>0$ to be specified later,
and define a function $v(\om;\th)$ on $\Om \times \Th$,
\be\label{eq:hull.haarsch}
v: (\om;\th) \mapsto \sum_{n=0}^\infty a_n \sum_{k=1}^{K_n} \th_{n,k} \ffi_{n,k}(\om),
\ee
which can be viewed as a family of functions $v_\th(\cdot)=v(\cdott;\th):\DT^\nu\to\DR$,
parameterized by $\th\in\Th$,
or as a particular case of a "random" series of functions, expanded over the given system of functions
$\ffink$ with "random" coefficients. It is to be emphasized that the orthogonality of the system
$\{\ffink\}$ is not important for our construction and results; for example, one could simply set
$\ffink = \one_{\Cnk}$, and this would even result in slightly simpler proofs.

We will call the expansions of the form \eqref{eq:hull.haarsch}
\emph{"haarsh" randelette expansions},  referring to Haar's (\emph{Haarsche}, in German)
wavelets and to the "harsh" nature of the
resulting potentials. Constructing a potential ``out of flat pieces''
is rather unusual in the framework of
the localization theory, where
all efforts  were usually made to avoid flatness of the potential.
Yet, with an infinite number of flat components $\thnk\,\ffink(\om)$, each modulated by its own
parameter $\thnk$, we proved earlier (cf. \cite{C01,C07a,C11c}) an analog of Wegner bound \cite{W81} for
the respective grand ensembles  $\{H(\om;\th), \om\in\Om, \th\in\Th\}$.

The extremely rapid decay of coefficients $a_n$ (the generation amplitudes),
making the series "lacunary", is required
for the proof of unimodality and of uniform decay of eigenfunctions. With generation amplitudes behaving like
$a_n \sim 2^{-bn}$, the tail series $\eps_{N+1}=\sum_{n\ge N+1} a_n$ is comparable to $a_N$, while
we need $\eps_{N+1} \ll |a_N|$.

We use the term "lacunary" for the following reason: instead of the series
$\sum_{n\ge0} a_n \cdot (\,\cdots\,)$
over all generations $n$, say, with $a_n = 2^{-bn^2}$,
we could consider a series of the form $\sum_{j=0}^\infty a_{n_j} \cdot (\,\cdots\,)$,
with $a_n = \eu^{-bn}$ and
a sequence $\{n_j\}$ growing fast enough; for example, $n_j = b j^2$. Such
series are usually called lacunary.

Building on the techniques from \cite{C11c}, we prove Anderson localization for
generic lacunary "haarsch"  potentials of  large amplitude,
under the mild assumptions \UPA (cf. \eqref{eq:cond.UPA}) and \DIV (cf. \eqref{eq:cond.DIV}).
In particular, our results imply uniform Anderson localization for a class of
quasi-periodic potentials with Diophantine frequencies.

\vskip3mm

Apparently, there is no hope to establish Anderson localization for a reasonably rich class
of quasi-periodic operators
without the assumption of strong disorder, even in one dimension,
as shows the well-known example of the Almost Mathieu
operator $H(\om) = \Delta + g \cos(n\alpha + \om)$ with Diophantine frequency
$\alpha$, featuring pure a.c. spectrum for $|g| < 2$. The approach based on the Lifshitz tails
asymptotics at ``extreme'' energies does not apply here.

\subsection{Main results}
\label{ssec:Main.res}

\btm\label{thm:Main}
Consider a family of lattice Schr\"{o}dinger operators in $\ell^2(\DZ^d)$,
$H(\om;\th) = \Delta + gV(x;\om;\th)$,
where  $V(x;\om;\th) = v(T^x\om;\th)$ with $v(\om;\th)$ given by the expansion \eqref{eq:hull.haarsch},
and the dynamical system $T^x$ satisfies conditions \UPA and \DIV  (cf. \eqref{eq:cond.UPA}, \eqref{eq:cond.DIV}) for some
$A, A', C_A, C_{A'} \in\DN^*$.

Then there exists $g_0 = g_0(C,A, C', A',d,\nu) \in(0,+\infty)$ such that for any
$g\ge  g_0$,  there exists a subset
$\Thinf(g) \subset \Th$ with
$\prth{\Thinf(g)} \ge 1 - \eu^{-c \ln^{1/2}g}$ and with the following property: if $\th\in\Thinf(g)$, then for \textbf{any} $\om\in\Om$:
\begin{enumerate}[\rm(A)]
  \item $H(\om;\th)$ has pure point spectrum;

  \item for any $x\in\DZ^d$, there is exactly one eigenfunction $\psi_x(\cdott;\om;\th)$ such that
\be\label{eq:psi.loc.center}
|\psi_x(x;\om;\th)|^2 > 1/2,
\ee
  i.e., $\psi_x$ has the ``localization center'' $x$, so the localization centers
  establish a bijection between the elements of the eigenbasis $\{\psi_x(\cdot;\om;\th)\}$
  and the lattice $\DZ^d$;

  \item for all $x\in\DZ^d$, the eigenfunctions $\psi_x$ decay uniformly away from their respective localization centers:
$$
\forall\,  y\in\DZ^d\;\; |\psi_x(y;\om;\th)| \le  \eu^{-m |y-x|}, \;
m = m(g, C, A) \tto{g\to+\infty} +\infty.
$$
\end{enumerate}
\etm

In Sect.~\ref{sec:DL.uniform}
we  establish \emph{uniform} pointwise dynamical localization for the operators $H(\om;\th)$
with $\th\in\Thinf(g)$ and any $\om\in\Om$ (cf. Theorem \ref{thm:DL.uniform}).

A direct analog of Theorem 5.2 proven in \cite{C11c} is the following

\btm
\label{thm:Minami.prob.Om.Th.1}
Fix a finite interval $I\subset\DR$. Then for some $B>0$, any $\om\in\Om$,
any integer $J\ge 1$ and some $C_J\in(0,+\infty)$
\be\label{eq:thm.Minami.prob.Om.Th}
\prth{ \th:\; \Tr \Pi_I( H_{\ball_L(0)}(\om;\th)) \ge J} \le  C_J\, L^{J B \ln L} |I|^J
\ee
and, denoting ${\DP^{\Omega\times\Theta}} := \DP\times\prthp$,
\be\label{eq:thm.Minami.prob.Om.Th.2}
 \promth{(\om,\th):\;  \Tr \Pi_I( H_{\ball_L(0)}(\om;\th)) \ge J} \le  C_J\, L^{J B \ln L} |I|^J.
\ee
\etm

Clearly, $J=1$ leads to a Wegner-type estimate. Theorem \ref{thm:Minami.prob.Om.Th.1}
is proved in Sect.~\ref{ssec:proof.Minami.Wegner}.

We also prove a variant of Theorem \ref{thm:Minami.prob.Om.Th.1} deterministic in $\om\in\Om$:

\btm\label{thm:Minami}
Consider a sequence $L_j = (L_0)^{2^j}\in\DN$, $L_0 > 1$.
Under the assumptions and with notations of Theorem \ref{thm:Main},
for any $g\ge g_0$, there exists a
subset $\ThinfM(g)\subset\Thinf(g)$ of measure $\prth{\Thinf(g)} \ge 1 - \eu^{-c_\rM \ln^{1/2}g}$,
and numbers $0 < B' < \tb <\infty$
such that for any $\th\in \ThinfM(g)$ and all $\om\in\Om$,
for any interval $I$ of length $|I|\le L_j^{-\tb \ln L_j}$
\be\label{eq:thm.Miniami}
\promth{ \Tr \Pi_I( H_{\ball_{L_j}(u)}(\om;\th)) \ge 2} \le  C'_2\, {L_j}^{ 2 B' \ln {L_j}} |I|^2.
\ee
\etm

Here the subscript "$\rM$" in $\ThinfM(g)$ refers to the Minami estimate.
The explicit values of the parameters $\tb, B'$ will be given in Sect.~\ref{sec:Minami}.
Using the Klein--Molchanov argument \cite{KM06}, we infer from \eqref{eq:thm.Miniami} the simplicity
of spectra of the operators $H(\om;\th)$ for all $\th\in\ThinfM(g)$ and \emph{every} $\om\in\Om$:

\btm\label{thm:simple.spectrum}
Under the assumptions and with notations of Theorem \ref{thm:Minami},
for any $g\ge g_0$, any $\th\in \ThinfM(g)$, and all $\om\in\Om$, $H(\om;\th)$
has \textbf{simple} pure point spectrum.
\etm

\section{Randelettes and separation bounds for the potential}
\label{sec:partitions}

\subsection{Relations between the key parameters}

In what follows, we often use parameters $A, C_A, A', B, b$ and some others; for the reader's
convenience, below are given the conditions they have to satisfy:
\be\label{eq:table.param}
\renewcommand{\arraystretch}{1.7}
\begin{tabular}{|l|l|}
  \hline
  $ b \ge \max\left( \frac{8d+4A+4A'}{10 A}, 2 \right) $ & $A \ln L_0 > |\ln C_A|+2\ln 2$
  \\

  \hline
   $ B = 800\, b A^2/ \ln 2 $ &  \,$\beta_0(g) = \eu^{- c_2 \ln^{1/2} g}$
   \\
  \hline
\end{tabular}
\ee

\subsection{Boundaries and partitions}

Given a lattice subset $\Lam\subsetneq \DZ^d$ with non-empty complement $\Lam^\rc$,
introduce its internal, external,
and the so-called edge boundary:
\be\label{eq:boundaries}
\bal
\pt^{-} \Lam &= \myset{x\in\Lam :\; \dist(x, \Lam^\rc) = 1},
\\
\pt^{+} \Lam &= \myset{x\in\Lam^\rc :\; \dist(x, \Lam) = 1} \equiv \pt^{-}\Lam^\rc,
\\
\pt \Lam &= \myset{(x,y)\in \pt^{-} \Lam \times \pt^{+} \Lam: |x-y|=1}.
\eal
\ee

Next, consider the phase space $\Om$ which we always assume to be the torus
$\DT^\nu$ of dimension $\nu \ge 1$: $\DT^\nu = \DR^\nu/\DZ^\nu \cong [0,1)^\nu$.
For each $n\ge 0$, we have introduced the family of
$K_n = 2^{\nu n}$ adjacent cubes $\Cnk$, $k=1, \ldots, K_n$,
of side length $2^{-n}$, and the functions $\ffink$ with $\supp \ffink = \Cnk$.

For every $n\ge 0$, the supports $\{\Cnk = \supp \ffi_{n,k}$,  $1 \le k \le K_n \}$
naturally define a partition of the phase space $\Om$:
$$
\cC_n = \myset{ C_{n,k}, 1\le k \le K_n  }.
$$
These partitions form a monotone sequence: $\cC_{n+1}  \prec \cC_n$, i.e., each element of $\cC_n$ is a
union of some elements of the partition $\cC_{n+1}$.

Given $n\ge 0$, for each $\om\in\Om$
we denote by $\hk_n(\om)$
the unique index such that
\be\label{eq:hk}
\om \in C_{n,\hk_n(\om)}.
\ee

\subsection{Piecewise-constant approximants of the hull}

For each $N\ge 0$, introduce the approximant of   $v(\om;\th)$ given by \eqref{eq:randelettes.1}:

\be\label{eq:v.xi.n.k}
v_N(\om;\th) = \sum_{n=0}^{N} \;a_n\; \sum_{k=1}^{K_n} \th_{n,k} \, \ffi_{n,k}(\om),
\ee
the truncated potential $V_{N}$ and the truncated Hamiltonian $H^{(N)}$:
\be\label{eq:V.N.H.N}
V_{N}(x;\om;\th) := v_{N}(T^x\om;\th), \;\;\;\;
H^{(N)} := \Delta + V_{N}.
\ee
With $b \ge 2$
(which follows from \eqref{eq:table.param}),
for any $N\ge 0$
we have
\be\label{eq:aN.decay.fast}
\bal
\sum_{n \ge N+1} a_n &= \sum_{n \ge N+1} 2^{-bn^2}
= 2^{-b(2N+1)} 2^{-bN^2} \sum_{i \ge 0} 2^{-b(N+i)^2 + b(N+1)^2}
\\
&\le 2^{-b(2N+1)} a_N \sum_{i \ge 0} 2^{-i} \le \half 2^{-2bN} a_N,
\eal
\ee
so the norm
$\| v - v_N \|_\infty := \sup_{\om\in\Om} \|  v - v_N \|_{L^\infty(\Th)}$
can be bounded as follows:
\be\label{eq:norm.v.vN}
\|  v - v_N \|_\infty \le \half 2^{-2bN} a_N.
\ee
Owing to \eqref{eq:aN.decay.fast}, the RHS is \textit{much smaller} than the width ($a_N$)
of the distribution of random coefficients $a_N\th_{N,k}$,
$1 \le k \le K_N$ (recall: $\th_{N,k} \sim \Unif[0,1]$).
Set
\be\label{tN.A.C}
\tnL = \tn(L,A,C_A) :=  1 + \left\lfloor \frac{4A \ln L - \ln (C_A/2)}{ \ln 2} \right\rfloor
\ee
and observe that, for $L$ large enough so $|\ln C_A| + 2\ln 2 < A \ln L$,
\be\label{eq:nt.A.C.less}
\bal
 \frac{3A }{ \ln 2} \ln L & < \;\;\,\tn(L)  & < &\; \frac{5A}{ \ln 2} \ln L,
\\
  L^{-5A} &<  \;2^{-\tn(L)} & < & \;\;L^{-3A} ,
\eal
\ee
Further, set
\be\label{eq:tN.vs.tn}
\tN(L) = \tn( L^4),
\ee
then we have
\be\label{eq:nt.tA}
\bal
%
   \tNL &= \tA \ln L, \;\; \tA = \tA(A, C_A)\in\left[\frac{12 A}{\ln 2}, \frac{20A}{\ln 2} \right]
   \subset \left[17 A, 29 A \right] .
\eal
\ee
and
\be\label{eq:Nt.A.C.less}
L^{-20 A} <  \;2^{-\tN(L)}  <  \;\;L^{-12 A} .
\ee
The condition $A \ln L_0 > |\ln C_A|+2\ln 2$
will be always assumed below (cf. \eqref{eq:table.param}).
Then for any $u\in\DZ^d$ and any $\om\in\Om$, all the points of the finite trajectory
$\{T^x\om, x\in\ball_{L^4}(u)\}$ are separated by the elements of the partition $\cC_{\tNL}$, since by
\UPA and
the first LHS inequality in \eqref{eq:nt.A.C.less}, we have
\be\label{nt.dist}
\half \dist_{\Om}(T^x\om, T^y\om) \ge \half C_A^{-1} \left( L^4\right)^{-A} >  2^{-\tNL}.
\ee

\ble\label{lem:LVB}
Under the assumptions \UPA and \DIV,
the bound \LVB holds true with $C''=1$ and $B = 800\, b A^2/\ln 2$.
\ele
\proof
Fix any integer $L\ge 1$ and let $\fB_L$ be the sigma-algebra generated by the random variables
$\{\thnk, n\ne \tNL, 1\le k \le K_n\}$. By \eqref{nt.dist},
all the points of the finite trajectory $\{T^x\om, x\in\ball_{L^4}(u)\}$ are separated by the
elements of the partition $\cC_{\tNL}$, so each value $v(T^x\om;\th)$ has the form
(we set for brevity $\widetilde{N} = \tNL$)
\be\label{eq:def.k.hat}
\bal
v(T^x\om;\th) &= \sum_{n \ne \tN} \; \sum_{k=1}^{K_n} a_n \thnk\,  \ffi_{n,k}(T^x\om)
+ \sum_{k=1}^{K_\tN} a_\tN \thnk \, \ffi_{\tN,k}(T^x\om)
\\
&= \zeta_\om(\th) + a_{\tN} \th_{\tN, \hk_{\tN}(T^x\om)}\, \fs_{\tN,\hk_\tN}(T^x\om), \quad
\fs_{\tN,\hk_\tN}(T^x\om) \in\{1, -1\},
\eal
\ee
where $\zeta_\om(\th)$ is $\fB_L$-measurable.
Since $\th_{\tN,\hk_\tN}\sim\Unif([0,1])$  and $\fs_{\tN,\hk_\tN}(T^x\om)=\pm 1$,
the second term in the above RHS has probability density bounded by
$$
a_n^{-1} = 2^{2b \tN^2} \le
\exp\left\{  \ln 2\cdot 2b \; \frac{ (20 A)^2 \ln^2 L}{ \ln^2 2} \right\}
=  L^{B \ln L}
$$
with
\be\label{eq:def.B}
B = 800\, b \,A^2 / \ln 2,
\ee
and it is independent of $\fB_L$.
This proves the claim.
\qedhere

\section{ Wegner-type bounds and spectral spacings}\label{WS_DoS}

We will use a sequence of integers (length scales) $L_j, j\ge 0$, defined as follows: given an
integer $L_0\ge 2$, we set
\be\label{eq:Lj}
L_{j} :=  L_{j-1}^{2} = (L_0)^{2^j}, \; j=1, 2, \ldots
\ee
A number of our formulae and estimates involve the cubes of size $L_j^{4}$; in view of the above definition,
$L_j^4 = (L_{j+1})^2 = L_{j+2}$, and the role of the quantities $L_j^4$ will become clear
at the finial stage of localization analysis, by the end of Sect. \ref{def:ell.q.dominated}.

In addition, in the proof of uniform exponential decay of eigenfunctions away
from their "localization centers", we will also use the length scale
\be\label{eq:Lj.minus.one}
L_{-1} =  0.
\ee
In Sect.~\ref{ssec:sep.fixed.omega.trunc},
we will introduce a function $g \mapsto L_0(g)$, providing for $g$ large enough
the value of the initial length scale suitable for the scale induction.
Here $g>0$ is the amplitude parameter in the potential $gV$ in \eqref{eq:DSO.intro}.
Specifically, we will show,
in the proof of Lemma \ref{lem:init.sep.V},
that it suffices to set, with some $c_1>0$,
\be\label{eq:def.L0.g}
L_0(g) = \left\lfloor \eu^{c_1 \ln^{1/2} g} \right\rfloor.
\ee
As a result, the length scales suitable for our scaling scheme become functions
of $g$: $L_j = L_j(g) = (L_0(g))^{2^j}$.
Next, given $g>0$,  set
\be\label{eq:def.delta.j}
\bal
\delta_j &= \delta_j(g) = \beta_j(g) \, a_{\tN(L_j)} \, ,
\\
\beta_j &= \beta_j(g) =  2^{-2 b \tN(L_j)} .
\eal
\ee
Here the function  $L \mapsto \tN(L)$ is defined in \eqref{tN.A.C}.
It follows that (cf. \eqref{eq:table.param})
\be\label{eq:beta.g}
\beta_0(g) \le \eu^{ - c_2 \ln^{1/2} g},
\ee
with some $c_2>0$ which will be specified later (in the proof of Lemma \ref{lem:init.sep.V}).
Observe that, owing to \eqref{eq:nt.A.C.less},
we have
\be\label{eq:bounds.on.delta.j}
\delta_j = 2^{-2b \tN(L_j)  } a_{\tN(L_j)}
<  L_j^{- (3A)^2 b \cdot 4\ln L_j} \le  L_0^{- C' 2^j \ln L_0 },
\ee
so that
$
\sum_{j\ge 0} \delta_j
%
< \infty.
$
Moreover,
$\sum_{j\ge 0} \delta_j \to 0$ as $L_0\to\infty$.

\subsection{The Wegner-type bound}
\label{ssec:Wegner}

As was said in Sect.~\ref{ssec:Main.res}, the particular case of Theorem \ref{thm:Minami.prob.Om.Th.1}
with $J=1$
gives a Wegner-type bound; it is non-uniform in the size of the cube $\ball_L(u)$, but
sufficient for the purposes of the scale induction  in Sect.~\ref{sec:MSA}.

\subsection{Parametric control of spectral spacings}

Consider a finite cube $\ball = \ball_L(u)\subset\DZ^d$ and the operator $H_\ball = \Delta_\ball + gV$.
If $g$ is large enough, then the values of the potential $\myset{V(x), x\in\ball}$ can be considered as
(satisfactory)
approximations to the eigenvalues $E^{\ball}_j$ of operator $H_\ball$, by virtue of the min-max principle.
In particular, if all the values of the potential in $\ball$ are distinct and $g$ is
large enough, then all spectral spacings $|E^{\ball}_i - E^{\ball}_j|$ of $H_\ball$ are bounded from below by
$C(V) g$.

A similar lower bound holds for all pairs of disjoint cubes
$\ball_\ell(u), \ball_\ell(v)$ inside a larger cube $\ball_L(w)$. Specifically,
if all the values
$\myset{V(x), x\in\ball_L(w)}$ are distinct and $g$ is large enough, then
$
|E^{\ball_\ell(u)}_i - E^{\ball_\ell(v)}_j| \ge C(V) g >0.
$
In other words,
the distance between the two spectra (as subsets of $\DR$) satisfies
$$
\dist\left[\Sigma\big(H_{\ball_\ell(u)}\big), \Sigma\big(H_{\ball_\ell(v)}\big) \right]
\ge Const(V) g >0.
$$
Here and below, we denote the spectrum of a finite-dimensional operator $H$ by $\Sigma(H)$.
In the case where $H = H_{\ball_L(u)}$, we will write for brevity $\Sigma(\ball_L(u))$.

However, it is clear that such a simple control of
inter-spectral spacings
is impossible at an
arbitrarily large scale, once $g$ is fixed.

In the traditional Multi-Scale Analysis of random operators, inter-spectral spacings are controlled in a probabilistic way, using the Wegner bound or its variants. The main \textit{raison d'\^{e}tre} of the auxiliary measurable space $\Th$ in the framework of the grand ensembles (cf. \cite{C11c}) is precisely to mimic,
to a certain extent, the Wegner-type bounds  and to assess the spectral
spacings for generic hulls $v:\Om\to \DR$.
%


Quite naturally, some hulls labeled by $\th\in\Th$ have to be excluded, essentially for the same reasons
that some samples of the IID random potentials have to be excluded in the proof of
localization: for example, setting all $\th_{n,k}=0$, we get $V(x;\om;\th)\equiv 0$,
hence the operator $H = \Delta$ with a.c. spectrum.

\vskip1mm

The above discussion suggests, and the analysis carried out below actually confirms
that, although the structure of probability (or, more generally, measure) space on the set of auxiliary
parameters $\thnk$ is a very convenient tool, it can be replaced by the structure of a metric space. The unwanted values of parameters  are covered by small balls, since the conditions required for a
successful application of the MSA procedure, in terms of the potentials and matrix elements
of the resolvents, have the form of inequalities. One particular advantage of the
probabilistic language
is the possibility to adapt
the conventional Wegner estimate in a straightforward way.

The role of Sect.~\ref{ssec:sep.spec.L0} and \ref{ssec:sep.spec.Lj} is to establish the crucial,
and quite unusual, property of the operators $H_{\ball_L(x)}(\om,\th)$: for any "good" $\th\in\Th$
(this notion will be made precise), and for any (not just $\DP$-a.e.) $\om\in\Om$, the spectra
of the operators $H_{\ball_L(x)}(\om,\th)$, $H_{\ball_L(y)}(\om,\th)$ in disjoint cubes with
$|x-y|\le L^4$ cannot be ``dangerously close'' to each other, so  the usual small denominators,
appearing in the perturbative scaling analysis, are \emph{never} excessively small.

Pictorially, \textbf{there are no resonances in our model}, exactly as in the Maryland
model and  its generalizations studied in  \cite{BLS83}.

\section{Separation of local spectra: initial scale}
\label{ssec:sep.spec.L0}

We work with the DSO $H_\ball$ in cubes $\ball=\ball_L(x)\subset\DZ^d$, with Dirichlet boundary conditions:
$H_\ball = \one_\ball H \one_\ball \upharpoonright \ell^2(\ball)$.
%
With $L=0$, $H_{\ball_0(x)}$ is the multiplication by $gV(x)$.

Given a function $V:\Lam\to\DR$ on a finite set $\Lam\subset\DZ^d$ (e.g., $\Lam=\ball$), let
\be\label{eq:def.sep.V}
\Sepb{V, \Lam} := \min\big\{ \,|V(x) - V(y)|,\; x,y\in\Lam, \, x\ne y \,\big\},
\ee
(here "Sep" stands for "separation [bound]"),
and for the operator $H_\Lam$,
\be\label{eq:def.sep.H}
\Sepb{ \Sigma(H_\Lam) } := \min\big\{ \,|E_i - E_j|,\; E_i, E_j\in\Sigma(H_\Lam), \, i\ne j \, \big\} .
\ee

\subsection{Separation bounds for fixed $\om\in\Om$}
\label{ssec:sep.fixed.omega.trunc}


Now we for\-mu\-late  the most important technical result of this paper.

\ble\label{lem:init.sep.V}
Assume the condition \UPA, with fixed the parameters $A, C_A$, and fix
the decay exponent $b>0$ in the definition of the sequence $\{a_n\}$ (cf. \eqref{eq:def.an.b}).
Then there exist constants $c_1, c_2>0$ with the following properties:

For all $g>0$ large enough there exists
an integer $L_0 =L_0(g) \ge \eu^{c_1 \ln^{1/2} g}$ (cf. \eqref{eq:def.L0.g}) and  a positive number
$\beta_0(g) \le \eu^{-c_2 \ln^{1/2}g}$ (cf. \eqref{eq:beta.g}) such that
for any $\om\in\Om$, any $u\in\DZ^d$, with $\delta_0 = \beta_0(g) a_{\tN(L_0)}$
(cf. \eqref{eq:def.delta.j}) one has
%
\be\label{eq:def.Th0.sep.V.tN}
\prth{ \Sep{ gV_{\tN(L_0)}(\cdot\,;\om;\th), \ball_{L_0^4}(u) } < 5 g \delta_0 }
\le C L_0^{8d } \beta_0(g)
\ee
and
\be\label{eq:def.Th0.sep.V}
\prth{ \Sep{ gV(\cdot\,;\om;\th), \ball_{L_0^4(u)} } < 4 g \delta_0 }
\le C L_0^{ 8d} \beta_0(g) .
\ee
Consequently, for any $m\ge 1$  there exists $g_* = g_*(m)\in(0, +\infty)$ such that
for all $g\ge g_*(m)$, the estimates \eqref{eq:def.Th0.sep.V.tN} and
\eqref{eq:def.Th0.sep.V} hold true, with $4g \delta_0 \ge 16 d \eu^{4m}$.

Equivalently, one can say that \eqref{eq:def.Th0.sep.V.tN}--\eqref{eq:def.Th0.sep.V} hold true
for sufficiently large $g>0$ with $4g \delta_0 \ge 16 d \eu^{4m(g)}$,
where $m(g)\to +\infty$ as $g\to+\infty$.
\ele

\proof
\textbf{1. Estimates for the truncated potential.}
Setting for brevity $\ball = \ball_{L_0}(u)$, we have for any $\tN\ge 1$ and $s>0$
\be\label{eq:mu.max.ball}
\bal
&\prth{ \min_{x\ne y\in \ball} |gV_\tN(x;\om;\th) - gV_\tN(y;\om;\th)| < gs} .
\\
& \le \half |\ball| \big( |\ball|-1 \big) \, \max_{x\ne y\in \ball} \,
   \prth{ |V_\tN(x;\om;\th) - V_\tN(y;\om;\th)| < s} .
\eal
\ee
Given $L_0\ge 2$, let $\tN = \tN(L_0)$, with $L\mapsto \tN(L)$  defined in \eqref{eq:tN.vs.tn}.
Fix $x\in\ball$, $\om\in\Om$; $V(x;\om;\th) \equiv v(T^x\om;\th)$
is a random variable of the form \eqref{eq:randelettes.1}, and (cf. \eqref{eq:def.k.hat})
\be\label{eq:sum.txi}
\bal
v_\tN(T^x\om;\th) &=
a_\tN \th_{\tN, \hk_\tN(x)} + \sum_{ n< \tN} a_n \th_{n, \hk_n(x)}
\\
&
=: a_\tN \th_{\tN, \hk_\tN(u)} + \hth_{\tN,x,\om}(\th) ,
\eal
\ee
where $\hth_{\tN,x,\om}$ is a sum of random variables (relative to $(\Th,\prthp)$), independent of
$\th_{\tN,\hk_\tN(u)}$.
By construction, $\th_{\tN,\hk_\tN(u)} \sim\Unif([0,1])$, so $a_\tN \th_{\tN,\hk_\tN(x)}$
admits the probability density bounded by $a_\tN^{-1}$, and so does the sum
$a_\tN \th_{\tN, \hk_\tN(u)} + \hth_{\tN,x,\om}(\th)$.

Similarly,
we decompose
\be\label{eq:sum.txi.y}
\bal
v_\tN(T^y\om;\th) &=
a_\tN \th_{\tN, \hk_\tN(y)} + \hth_{\tN,y,\om}(\th) .
\eal
\ee
By definition of $\tN = \tN(L_0) = \tn(L_0^4)$, the elements of the partition $\cC_\tN$ separate the points
$T^x\om$ and $T^y\om$,
for all $x,y\in\ball_{L_0^4}(0)$, $x\ne y$,
thus $\hk_\tN(x) \ne \hk_\tN(y)$, and $\th_{\tN,\hk_\tN(x)}$
is independent of $\th_{\tN,\hk_\tN(y)}$.

Denote
$\BX(\th) = \th_{\tN, \hk_\tN(x)}$, $\BY(\th) =\th_{\tN, \hk_\tN(y)}$
($\om$ is fixed and omitted); then
$$
\bal
\prth{ |v_\tN(T^x;\om;\th) - v_\tN(T^y;\om;\th)| \le t \Bcond \fB_{L^4_0} }
= \prth{ |\BX - \BY - \BZ | \le a_\tN^{-1} t \Bcond \fB_{L^4_0} }
\eal
$$
where the random variable (we omit again its parameter $\om$ which is fixed)
$$
\BZ(\th) = a_{\tN}^{-1}\left(\hth_{\tN,y,\om}(\th) - \hth_{\tN,x,\om}(\th) \right)
$$
is $\fB_{L^4_0}$-measurable, so we have, $\prthp$-a.s.,
$$
\bal
\prth{ |\BX - \BY - \BZ| \le a_{\tN}^{-1} t \Bcond \fB_{L^4_0} }
&\le \sup_{s\in\DR}\; \prth{ |\BX - \BY - s| \le a_\tN^{-1} t \Bcond \fB_{L^4_0} }
\\
& \le 2  a_\tN^{-1} t,
\eal
$$
since $\BX \sim \Unif[0,1]$ and $\BY$ is independent of $\BX$,
so $\BX - \BY$ has density $\le 1$. Thus
$$
\bal
&\prth{ |v_\tN(T^x;\om;\th) - v_\tN(T^y;\om;\th)| \le t }
\\
&\quad
= \esmth{ \prth{ |v_\tN(T^x;\om;\th) - v_\tN(T^x;\om;\th)| \le t \cond \fB_{L^4_0} } }
\\
& \quad \le 2  a_\tN^{-1} t .
\eal
$$
Recalling \eqref{eq:mu.max.ball}, we conclude that
\be\label{eq:prthp.gs}
\bal
\prth{ \min_{x\ne y\in \ball} |gV_\tN(x;\om;\th) - gV_\tN(y;\om;\th)| < g s}
\le C L_0^{8d} a_\tN^{-1} s.
\eal
\ee

\vskip1mm
\noindent
\textbf{2. Perturbation estimates.}
Let $\beta>0$ (a suitable value of $\beta$ will be specified later) and
$$
s = 5 \beta\, a_\tN .
$$
Then we infer from \eqref{eq:prthp.gs} that
\be\label{eq:prth.Th.0.om}
\bal
\prth{ \Sep{ gV_\tN, \ball_{L_0^4}}  < 5g \beta\,  a_\tN }
\le C' \, L_0^{8d} \, \beta .
\eal
\ee
Let
\be\label{eq:def.event.Thminus}
\Thminus(g,\om) :=
\left\{
\th\in\Th:\; \Sep{ gV_\tN, \ball_{L_0^4}}  \ge 5g \beta  a_\tN
\right\} ,
\ee
then by \eqref{eq:prth.Th.0.om}, we have
\be\label{eq:prthp.Thzero}
\prth{ \Thminus(g,\om) } \ge 1 - C'  \, L_0^{8d}  \beta.
\ee
On the other hand,
$\| gV - gV_\tN\|_\infty \le \half g 2^{- 2b\tN} a_\tN$ (cf. \eqref{eq:norm.v.vN}).
Now set
\be\label{eq:beta.g.again}
\beta = \beta_0(L_0) :=  2^{-2b\tN(L_0)} ,
\ee
then
$$
\| gV - gV_\tN\|_\infty \le \half g 2^{-2 b \tN} a_\tN = \half g \beta  a_\tN = \half g\delta_0.
$$
Thus for any $\th\in\Thminus(g,\om)$,  we have, by triangle inequality,
\be\label{eq:sep.bound.nontruncated.V}
\bal
\Sep{ gV, \ball_{L_0^4}} &\ge \Sep{ gV_\tN, \ball_{L_0^4}} - 2 \| gV - gV_\tN\|_\infty
\\
%
& \ge 5 g \delta_0  - 2 \cdot \half g \delta_0
\\
&
= 4 g \delta_0 = 4 g\,  2^{-2b\tN } a_{\tN} .
\eal
\ee

Further, we need the quantity $\Sep{ gV, \ball_{2 L_0^4}}$ to be large, viz.
\be\label{eq:Sep.5.e.m}
\Sep{ gV, \ball_{L_0^4}} \ge 16 d\, \eu^{4m}, \; m \ge 1.
\ee
On the account of the lower bound \eqref{eq:sep.bound.nontruncated.V}, it suffices that
$$
4 g\,  2^{-2b\tN } a_{\tN} \ge 16 d\, \eu^{4m}, \;\; \text{ where } a_{\tN(L_0)} = 2^{-b \tN^2(L_0)}.
$$
Consequently, given the numbers $g>0$, $m\ge 1$, we set
\be\label{eq:choice.L_0.g}
\bal
L_0(g) = L_0(g,m) &:=
\max \left\{L_0\in \DN:\; 4 d\, \eu^{4m} \, 2^{b \tN^2(L_0) + 2b\tN(L_0)  } \le g \right\},
\\
\beta_0(g) = \beta_0(g,m) &:= \beta(L_0(g,m)).
\eal
\ee
Then it is readily seen that, for any fixed $m$,
\be\label{eq:beta.L0}
\lim_{g\to\infty} L_0(g) = +\infty, \;\;
\lim_{g\to+\infty} \beta_0(g) = 0.
\ee
Indeed, recall that
$\tN(L_0) \le 29 A \ln L_0$
(cf. \eqref{eq:nt.tA}; ; since $\tN^2 > 2\tN$ for $\tN>1$,
we have
$$
b \tN^2(L_0) + 2b\tN(L_0) \le 2 b \tN^2(L_0) \le  2 \cdot (29 A \ln L_0 )^2
$$
%
Therefore, are  admissible in \eqref{eq:choice.L_0.g}
the integers $L_0$ such that
$$
\ln^2 L_0 \le \frac{ \ln g - \ln \big( 4d \eu^{4m} \big)  }{ 2b (29 A)^2}.
$$
For $g$ large enough, so $\frac{1}{2} \ln g\ge  \ln (4d \eu^{4m}) $,
the above RHS is bigger than
$
\frac{\ln g }{ 4 b (29 A)^2 },
$
thus for such $g>0$,
the maximum $L_0(g)$ in \eqref{eq:choice.L_0.g}
satisfies the lower bound
\be\label{eq:cond.g.L0.1}
\ln L_0(g) \ge c_1 \ln^{1/2} g, \; \; c_1 = c_1(A,b) := \frac{1}{58 A \sqrt{b}}  \, .
\ee
One can transform it into a formal definition, setting
$ L_0(g) := \lfloor \eu^{c_1 \ln^{1/2} g }\rfloor$.

The quantity $\beta(L_0)$, with $L_0 = L_0(g)$, becomes a function of $g$, and we have
\be\label{eq:lamma.sep.params.beta0}
\bal
\beta_0(g) =
\beta(L_0(g)) = 2^{-b \tN(L_0(g))} = 2^{-4 b  \tA \ln L_0(g)}
\le \eu^{ -c_2 \ln^{1/2} g } ,
\eal
\ee
with $c_2 = c_2(b, A, C_A) \ge 68 A b$ (recall that by  \eqref{eq:nt.tA}, $\tA \ge 17 A$).

Collecting \eqref{eq:prth.Th.0.om}, \eqref{eq:def.event.Thminus}, \eqref{eq:prthp.Thzero}
and \eqref{eq:Nt.A.C.less},
we obtain
\be\label{eq:lemma.sep.finial.bound}
\bal
\prth{ \Thminus(g, \om) } \equiv
\prth{ \Sep{ gV_\tN, \ball_{L_0^4}} \ge 5g \delta_0 }
& \ge 1 - C \, L_0^{8d} \, \beta_0(g)
\\
& \ge 1 - C \, L_0^{- 12 b A +8d} .
\eal
\ee
Since the inequality $\Sep{ gV_\tN, \ball_{L_0^4}}  \ge 5g \delta_0$
implies $\Sep{ gV, \ball_{L_0^4}}  \ge 4g \delta_0$ (cf. \eqref{eq:sep.bound.nontruncated.V}),
both asserted bounds
\eqref{eq:def.Th0.sep.V.tN}--\eqref{eq:def.Th0.sep.V}
follow from \eqref{eq:lemma.sep.finial.bound}.

For any $m\ge 1$, there is a sufficiently large $g_*(m)>0$ such that
for $g\ge g_*(m)$,
\be\label{eq:Sep.5.e.m.rem}
\Sep{ gV, \ball_{L_0^4}} \ge 4 g \delta_0 \ge 16 d\, \eu^{4m} ,
\ee
since
$$
 g \delta_0(g) \ge g \eu^{-c \ln^{1/2} g} = \eu^{\ln g \left(1 -c \ln^{-1/2} g\right)}
 = g^{ 1 - o(1)}, \;\; \text{ as } g\to\infty.
$$
One can start with $m\ge 1$ and find an appropriate lower threshold $g_*(m)$  for $g$,
or start with $g>0$ large enough and define
\be\label{eq:def.m(g)}
m = m(g) := \frac{1}{4} \ln \frac{g\delta_0(g)}{4 d}
\tto{g\to +\infty}+\infty.
\ee
\qedhere

\vskip3mm

\noindent
$\blacklozenge$ We stress that, albeit the subsets $\Thminus(g,\om)\subset\Th$ depend upon $\om\in\Om$,
the lower bound
\eqref{eq:lemma.sep.finial.bound}
on $\prth{\Thminus(g,\om)}$
is uniform in $\om$.

\subsection{Separation of finite trajectories}
\label{ssec:sep.traj}

Introduce some geometrical objects related to the length scales $L_j$, $j\ge 0$.
First, let
\be\label{eq:def.R.j}
R_j = \frac{1}{6} C_A^{-1} L_j^{-4A}
\ee
(recall that $A, C_A\in\DN^*$),
and cover the torus $\Om$ redundantly by the union of $N_{R_j}:= (R_j)^{-\nu }$ cubes $Q_{3R_j}(\om_i)$, $i\in[[1,N_{R_j}]]$, of radius $3R_j$ and with centers of the form
$$
\om_i = \left[ l_1 R_j, \ldots, l_\nu  R_j  \right),
\; l_1, \ldots, l_\nu \in [[0, (2R_j)^{-1}-1]].
$$
The order of numbering can be arbitrary.
Next, decompose each cube $Q_{3R_j}(\om_i)$ into a union of $3^\nu $ neighboring sub-cubes
$Q'_{R_j}(\om'_{i,k})$ of radius $R_j$, which we number starting from the central sub-cube,
$Q'_{R_j}(\om'_{i,1})$. Observe  that the collection of all central sub-cubes $Q'_{i,1}(R_j)$
covers the torus $\Om$, and $\om'_{i,1} \equiv \om_i$.

Similarly, cover the torus $\Om$ by adjacent cubes $Q_{r_j}(\om''_i)$ of
radius
\be\label{eq:def.rj}
r_j = C^{-1}_{A'}L_j^{-4A'} R_j = \big(6 C_{A'} C_A\, L_j^{4A+4A'} \big)^{-1} .
\ee

\ble\label{lem:dist.r.R}
{\rm(See Fig. 1.)}
Fix $j\ge 0$ and consider $\ball_{L_j^4}(0)\subset\DZ^d$.
Fix any
point
$z\in\ball_{L^4_j}(0)$ and a cube $Q_{r_j}(\om''_i)$. If
$T^z\om''_i\in Q'_{R_j}(\om'_{\icirc,1})$ for some $i_\circ = i_\circ(i,z)$, then
\be\label{eq:lem.dist.r.R}
T^z \big(  Q_{r_j}(\om''_i) \big) \subset  Q_{3R_j}(\om_{i_\circ}) \equiv Q_{3R_j}(\om'_{i_\circ,1}).
\ee
\ele

\begin{center}
\begin{figure}
\begin{tikzpicture}
\begin{scope}[scale=0.9]
\clip (-2.1,-1) rectangle ++(12.5,6.3);

\fill[lightgray] (2.5,1.5) rectangle (3,2);
\fill (2.75,1.75) circle (0.07);

\foreach \i in {0, 0.5, 1, 1.5, 2, 2.5, 3, 3.5}
\draw (\i, -0.5) -- (\i, 4);

\foreach \i in {0, 0.5, 1, 1.5, 2, 2.5, 3, 3.5}
\draw (-0.5, \i) -- (4, \i);

\fill[color=gray!50!white!40] (7.33, 2.33) rectangle ++(2.33, 2.33);

\foreach \i in {7, 8, 9, 10 }
\draw (\i, 0.0+2) -- (\i, 4+1);

\foreach \i in {0, 1, 2, 3}
\draw (7, \i+2) -- (10, \i+2);

\draw[line width = 3] (7, 2) rectangle (10, 5);

\fill (8.51, 3.49) circle (0.07);
\node at (5.7, 4.75) (omizero) {$\om_{i_\circ}\equiv \om'_{i_\circ, 1}$};
\draw[->,bend right=15] (omizero.south) to (8.37, 3.49);

\fill (8.83, 3.33) circle (0.12);
\node at (2.8, 1.75) (om) {};
\draw[->,bend right=15, line width = 1.2] (om.east) to (8.71, 3.25);

\node at (4.9, 2.2) (Tx) {$T^z$};

\node at (5.3, 0) (omi) {$\om''_{i}$};
\draw[->, bend left=30] (omi.west) to (2.81, 1.64);

\node at (9.3, 1.5) (QR) {$Q_{3R_j}(\om_{i_\circ})$};

\node at (6.5, 1.2) (QRprime) {$Q'_{R_j}(\om'_{i_\circ,1})$};
\draw[->,bend right=25] (QRprime.east) to (8.7, 2.93);

\draw[line width = 1.2] (8,3) rectangle (9,4);

\node at (5.5,3.3) (Qr) {$Q_{r_j}(\om''_i)$};
\draw[->,bend right=45] (Qr.west) to (2.8,2.05);

\end{scope}
\end{tikzpicture}

\caption{\emph{Example for Lemma \ref{lem:dist.r.R}}.
If $T^z \om''_{i}$ hits the central sub-cube $Q'_{R_j}(\om'_{i_\circ,1})$,
then, owing to the condition \DIV,
the image by $T^z$ of the entire cube
$Q_{r_j}(\om''_{i})$ (gray) must be contained in a subset of the larger, concentric
cube $Q_{3R_j}(\om'_{i_\circ,1})$,
represented by the light-gray area.
}
\end{figure}
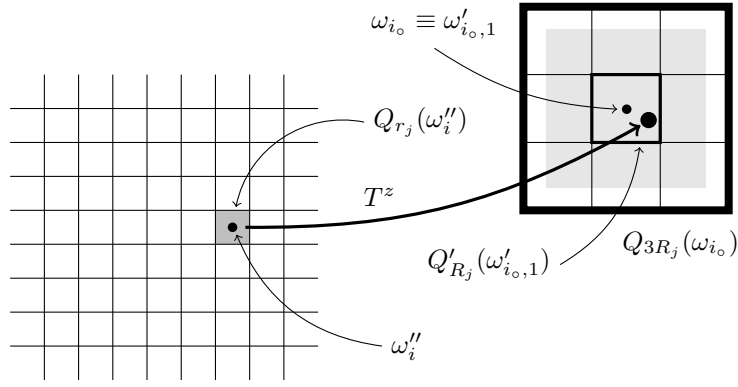
\end{center}

\proof
For any $\om\in Q_{r_j}(\om''_i)$, we have $\dist(\om''_i, \om)\le r_j$, thus
by \DIV and \eqref{eq:def.rj},
\be\label{eq:dist.r.R.1}
\dist( T^z \om''_i, T^z\om) \le C_{A'}(L_j^4)^{A'} \dist(\om''_i, \om) \le C_{A'} L_j^{4A'} r_j
=  R_j.
\ee
By assumption,
\be\label{eq:dist.r.R.2}
\dist(T^z \om''_i, \om_{i_\circ}) \equiv
\dist(T^z \om''_i, \om'_{i_\circ,1}) \le R_j,
\ee
therefore, by \eqref{eq:dist.r.R.1} and \eqref{eq:dist.r.R.2},
$$
\bal
\dist(T^z \om, \om_{i_\circ}) &\le
\dist(T^z \om, T^z\om'_{i}) + \dist(T^z\om'_{i}, \om'_{i_\circ})
\le R_j + R_j < 3R_j,
\eal
$$
yielding the assertion \eqref{eq:lem.dist.r.R}.
\qedhere

\vskip3mm

For each $j\ge -1$, define the integers
\be\label{eq:bound.on.Nj}
\tN_j=
\begin{cases}
\tN(L_j), & j\ge 0
\\
\tN(L_0), & j = -1
\end{cases},
\qquad
\cL_j=
\begin{cases}
 \big(12 C_A C_{A'}\big)^\nu L_j^{4\nu(A+A')}, & j\ge 0
\\
\big(12 C_A C_{A'}\big)^\nu L_0^{4\nu(A+A')}, & j = -1
\end{cases},
\ee
with $L  \mapsto \tN(L) = O(\ln L)$ defined in \eqref{eq:tN.vs.tn}.

\bco\label{cor:finite.cover.cubes}
Fix any integer $j\ge 0$.
There exists a finite collection of points,
$\cT_j = \{\taujl, \, 1\le l \le \cL'_j \le \cL_j\}$,
and a measurable partition of $\Om = \DT^\nu$,
$\cP_j = \big\{ \rP_{j,l} \ni \taujl, \; 1\le l \le \cL'_j  \big\}$, such that
\par\noindent
$\bullet$ any cube $Q_{r_j}(\om''_i)$ is covered by at most $2^\nu$ elements
of $\cP_j$;

\par\noindent
$\bullet$   for every $z\in\ball_{L_j^4}(0)$, the image
$T^z \rP_{j,l}$ is covered by exactly one element of the partition $\cC_{\tN(L_j)+1}$.

\eco
\proof
For notational brevity, let $n=\tN(L_j)$.
Fix a cube $Q_{r_j}(\om''_i)\subset\Om$ and any $z\in\ball_{L_j^4}(0)$.
Consider the image $T^z Q_{r_j}(\om''_i)$. By Lemma \ref{lem:dist.r.R}, it is covered by
one cube $Q_{3R_j}(\om_\icirc)$, with some $\icirc = \icirc(n,i,j)$.
Since by \eqref{eq:def.R.j} and \eqref{nt.dist} we have
$$
\diam Q_{3R_j}(\om_\icirc)=6R_j = C_A^{-1} (L_j^4)^{-A} < 2^{-\tN(L_j)-1} = 2^{-n-1},
$$
the the image $T^z Q_{r_j}(\om''_i)$  is covered by at most
$2^\nu$ adjacent cubes of side length $2^{-n-1}$ -- the elements of the partition $\cC_{n+1}$,
which are sub-cubes of $Q_{3R_j}(\om_\icirc)$.
Following the notation introduced in
Sect.~\ref{ssec:lacunary.radelettes} (cf. \eqref{eq:Cbullnk}),
denote these cubes by $C_{n,k_l;i_l}$, $l=1, \ldots, 2^\nu$. (Recall that,
by definition, for each pair $(n,k)$, the cube $\Cnk=\supp\ffink$ is partitioned
in to the sub-cubes $C_{n,k;i}$, $i=1, \ldots, 2^\nu$, of side length $2^{-n-1}$,
on which the Haar's wavelet $\ffink$ takes constant value $\pm 1$.)

Now the required partition $\cP_j$ can be formed by taking all the non-empty intersections
of the form
$T^{-z}C_{n,k_l;i_l} \cap Q_{r_j}(\om''_i)$,
$l=1, \ldots, 2^\nu$, for all $i$.
For the collection $\cT_j$, it suffices to pick exactly one point from each set $\rP_{j,l}$,
and denote it by $\taujl$. Since the number of cubes $Q_{r_j}(\om''_{i})\subset\DT^\nu$
of diameter
$r_j$
is bounded by
$r_j^{-\nu} = \big(6 C_A C_{A'} L_j^{4A+4A'}\big)^\nu $,
we have $\card \cT_j\le (12 C_A C_{A'})^\nu  L_j^{4\nu(A+A')}=\cL_j$, as asserted.
\qedhere

\vskip3mm

Now define the operator-valued mappings
\be\label{eq:def.fh.j.positive}
\fh^{\Nj}_{j,\th}:
\om \mapsto
\begin{cases}
H^{(\Nj)}_{\ball_{ L_j^4}(0)}(\om;\th)\upharpoonright \ell^2(\ball_{ L_j^4}(0)),
   & j\ge 0 \\
gV_{\tN(L_0)}(\cdot\,;\om;\th) \upharpoonright \ell^2(\ball_{ L_0^4}(0)), & j=-1
\end{cases} .
\ee
In the above formula, $gV_{\tN(L_0)}(\cdot\,;\om;\th)$ is the truncated potential
on $\ball_{ L_0^4}(0)$, identified with the multiplication operator by this potential.

\ble\label{lem:loc.const.H}
Fix any $j\ge -1$.
For any fixed $\th\in\Th$, the mapping $\fh^{\Nj}_{j,\th}$,
defined in \eqref{eq:def.fh.j.positive},
is piecewise-constant on $\Om$.

More precisely, let the collection $\cT_j$ and the partition $\cP_j$ be defined as in
Corollary \ref{cor:finite.cover.cubes}.
Then $\fh^{\Nj}_{j,\th}$ is constant on each element $\rP_{j,l}$ of $\cP_j$. Thus
the operator-valued function $\fh^{\Nj}_{j,\th}$ takes on $\Om$ only a finite number of values,
\be
\fh^{\Nj}_{j,\th}(\taujl), \;\; 1 \le l \le \cL'_j \le \cL_j.
\ee
\ele

\proof
Fix $j\ge -1$ and let $\tN_j$ be given by \eqref{eq:bound.on.Nj}.
%
%
By Corollary \ref{cor:finite.cover.cubes},
the truncated hull $v_{\tN_j}$ is constant on each element of the partition $\cP_j$,
and so is, therefore, the function
$\fh^{\Nj}_{j,\th}$, since the
kinetic energy
operator $\Delta$
(present in $H^{(\Nj)}_{\ball_{ L_j^4}(0)}(\om;\th)$ for $j\ge 0$)
is constant in
$\om\in\Om$ (and in $\th\in\Th$). This proves the claim.
\qedhere

\vskip2mm

\subsection{Separation bounds uniform in $\om\in\Om$}

\ble\label{lem:bound.Sep.V.uniform.omega}
For all  $g>0$  large enough, there is a measurable subset $\Th^{(-1)}(g)\subset\Th$ with
\be\label{eq:bound.lem.Sep.V.uniform.omega}
\prth{ \Thminus(g) } \ge 1 - C L_0^{ - 12 b A + 8d + 4\nu(A+A')}
\ee
such that for any $\th\in\Th^{(-1)}(g)$ and every $\om\in\Om$, one has
$$
\Sep{gV,L_0^{4}} \ge 4g \delta_0 .
$$
\ele

\proof

Consider the sets
$$
\Th \supset \rP_{-1,l} \ni \tauminusl, \;\; l=1, \ldots,
  \cL'_{-1} \le \cL_{-1},
$$
introduced in the Sect.~\ref{ssec:sep.traj}. By Lemma \ref{lem:loc.const.H}, the function
$$
\fh^{\Nminus}_{-1,\th}: \om \mapsto  gV_{\tN(L_0)}(\cdot\,;\om;\th)
   \upharpoonright \ball_{L_0^4}(0)
$$
is constant on each $\rP_{-1,l}$, so if the required separation bound holds true
for each phase point $\tauminusl$, $l\in[1, \cL_{-1}]$, then it also holds for every $\om\in\Om$.
By Lemma \ref{lem:init.sep.V} and Eqn. \eqref{eq:lemma.sep.finial.bound},
which apply to any $\om\in\Om$, including of course $\tauminusl$,
$$
\prth{ \Sep{gV(\cdot;\tauminusl;\th), L_0^4} < 4 g \delta_0(g)  }
\le C L_0^{- 12 b A + 8d}.
$$
By \eqref{eq:bound.on.Nj}, $\cL'_{-1} \le \cL_{-1}= \Const L_0^{4\nu(A+A')}$, yielding
\eqref{eq:bound.lem.Sep.V.uniform.omega}.
\qedhere

\subsection{Uniform separation bounds for spectra: The initial scale}

\bde\label{DefNR} Given $E\in\DR$ and a DSO $H_{\ballj(x)}$,
the cube $\ballj(x)$ is
called $E$-non-resonant (\ENR) if  the following bound holds:
$$
\dist\Big[ \Sigma\big(H_{\ballj(x)}\big), E \Big] \ge
  g \delta_j .
$$
Otherwise, it is called $E$-resonant ($E$-R).
\ede

\bde
Given a DSO $H_{\ball_L(x)}$ in a cube $\ball_L(x)$,
we say that $\ball_L(x)$ is $(E,m)$-non-singular (\EmNS), with $E\in\DR$, $m>0$,
if for any $y\in\pt^- \ball_L(x)$ (\emph{cf. the definition of $\pt^-$ in \eqref{eq:boundaries}})
\be\label{eq:GF.L0.2}
|G_{\ball_L(x)}(x,y;E)| \le
\left\{
  \begin{array}{ll}
   (3L)^{-d}\, \eu^{-\gamma(m,L)}, & \hbox{if $L\ge 1,$} \\
   (2d)^{-1} \eu^{-\gamma(m,L)}, & \hbox{if $L=0,$}
  \end{array}
\right.
\ee
where
\be\label{eq:def.gamma}
\gamma(m,L) :=
\left\{
  \begin{array}{ll}
   m(1 + L^{-1/8})L , & \hbox{if $L\ge 1$,} \\
   2m, & \hbox{if $L=0$.}
  \end{array}
\right.
\ee
Otherwise, $\ball_L(u)$ is called $(E,m)$-singular (\EmS).
\ede

\ble\label{lem:sep.L0}
Let the subset $\Thminus(g)\subset\Th$ be defined as in Lemma \ref{lem:bound.Sep.V.uniform.omega}.
For any $m \ge 1$, there exist an integer $L_0 = L_0(m)\ge 2$ and a real number $g_*(m)>0$
such that for $g \ge g_*(m)$ and for any $\th\in\Thminus(g)$,
any $\om\in\Om$, any $u\in\DZ$ and any $E\in\DR$, there is at most one
single-site cube $\{x\} = \ball_{0}(x)\subset \ball_{L_0^4}(u)$ which is \EmS.
\ele
\proof
It follows from
the definition of the subset
$\Thminus(g)$ (cf. Lemma \ref{lem:bound.Sep.V.uniform.omega} and
\eqref{eq:bound.lem.Sep.V.uniform.omega}
%
%
that, for an arbitrarily large $m\ge 1$
and all $g>0$ large enough, so that $m(g) \ge m$ with $m(g)$ defined as in
\eqref{eq:def.m(g)},
for all $x,y\in\ball_{L_0^4}(0)$ with $x\ne y$,
$$
|gV(x;\om;\th) - gV(y;\om;\th) | \ge 4 g \delta_0 \ge 16 d\eu^{4 m},
$$
thus there is no pair of points $x,y\in\ball_{L_0^4}(u)$, $x\ne y$, such that
\be\label{eq:GF.L0.1}
\max \big\{\,  |gV(x;\om;\th - E|, \, |gV(y;\om;\th) - E| \, \big\} < 2 g \delta_0,
\ee
Given any $x\ne y$ in $\ball_{L_0^4}(u)$, for at least one point $z\in\{x,y\}$, we have
$$
\bal
\| G_{\ball_0(z)}(E)\| = \big| (gV(z;\om;\th) - E)^{-1}   \big|
\le (2g \delta_0)^{-1}
& \le (8 d)^{-1} \eu^{-4m}
\\
& < (2d)^{-1} \eu^{ -\gamma(m,0)},
\eal
$$
yielding the \EmNS property of $\ball_{0}(z)$. Hence no pair of distinct single-site cubes
$\ball_0(x)$, $\ball_0(y)\subset \ball_{L_0^4}(u)$ can be \EmS for the same value of $E\in\DR$.
\qedhere

\vskip2mm


Decay of the Green functions in the balls of radius $L_0>0$
can be assessed  with the help of a variant of the Combes--Thomas estimate
(cf. \cite{CT73}, \cite{K08}) adapted to large spectral gaps.

\bpr[Cf. \cite{CS13}*{Theorem 2.3.4}]\label{prop:single.E.S.L0}
Suppose that for some $E\in\DR$, one has
$\dist(E, \sigma(H_{\ball_L(u)})) \ge \eta > 4d$. Then for any $x,y\in\ball_L(u)$
\be\label{eq:lem.CT}
| G_{\ball_L(u)}(x,y;E)| \le  2 \eta^{-1} \eu^{ - \mu |x-y|} <  \eu^{ - \mu |x-y|}
\ee
with
\be\label{eq:lem.CT.m}
\mu = \half \ln \frac{\eta}{4d} .
\ee
Consequently, for large $g>0$,
any cube $\ball_{L_0(g)}(u)$ which is \ENR is also \EmNS.
\epr
\proof
The first assertion \eqref{eq:lem.CT}
is proved in \cite{CS13}. If  the cube $\ball_{L_0}(u)$  is \ENR,
then $\dist(E, \Sigma(H_{\ball_L(u)})) \ge \eta >0$, with $\eta = g\delta_0 \ge 4d\eu^{4m} > 4d$,
so \eqref{eq:lem.CT} implies
\be
| G_{\ball_L(u)}(x,y;E)| \le  \eu^{ - \mu |x-y|} ,
\ee
where
$$
\mu = \half \ln \frac{ 4d \eu^{4m}}{4d} = 2m = \gamma(m,L_0) + m\big( L_0 - L_0^{7/8} \big).
$$
For $L_0 \ge 3$, one has $L_0 - L_0^{7/8} > \half L_0$; for $g>0$ large and $L_0 = L_0(g)$,
the latter condition is
fulfilled. Further, for $|x-y|=L_0$ and $g$ large enough,
we have
$$
\eu^{ - \mu |x-y|} \le  \eu^{-\gamma(m,L_0)} \eu^{-mL_0/2} \le (3L_0)^{-d} \eu^{-\gamma(m,L_0)},
$$
thus $\ball_{L_0}(u)$ is \EmNS.
\qedhere


\ble
Let be given real numbers $m \ge 1$
and sufficiently large $g>0$, so that $m(g)\ge m$ (with $m(g)$ defined in \eqref{eq:def.m(g)}).
Let  $L_0 = L_0(g)$, $\delta_0 = \delta_0(g)$.
Then for any $(\om,\th)\in\Om\times\Thminus(g)$, any $u\in\DZ$ and any $E\in\DR$,
there is no pair of disjoint \EmS
cubes $\ball_{L_0}(x), \ball_{L_0}(y)\subset \ball_{L_0^4}(u)$.
\ele

\proof
Consider any cube
$\ball_{L_0^4}(u)$. Suppose that some cube $\ball_{L_0}(x)\subset \ball_{L_0^4}(u)$
is \EmS; then it must be \ER, for otherwise it would be \EmNS, by  Proposition \ref{prop:single.E.S.L0}.
Therefore,
\be\label{eq:dist.E.V.x0}
\exists\, x_0\in\ball_{L_0}(x)\;\; |gV(x_0;\th;\om) - E| \le g\delta_0.
\ee
Consider any cube $\ball_{L_0}(y)\subset \ball_{L_0^4}(u)$ disjoint from
$\ball_{L_0}(x)$. For any $\th\in\Thminus(g)$ and any $\om\in\Om$,
$\Sep{V, \ball_{L_0^4}(u)} \ge 4 g\delta_0$, thus
\be\label{eq:dist.V.x0.V.z}
\min_{z\in\ball_{L_0}(y)} \left| gV(x_0;\th;\om) - gV(z;\th;\om) \right| \ge 4 g\delta_0,
\ee
so by the triangle inequality combined with \eqref{eq:dist.E.V.x0} and \eqref{eq:dist.V.x0.V.z},
$$
\min_{z\in\ball_{L_0}(y)} \left| gV(z;\th;\om) - E \right| \ge 4 g\delta_0 - g\delta_0 = 3g\delta_0.
$$
Further, by the min-max principle, considering $H$ as a perturbation of $gV$ by $H_0=\Delta$, we have
$$
\dist\left( \Sigma(H_{\ball_{L_0}(y)}), E \right) \ge
\dist\left( \Sigma\big( (gV)_{\ball_{L_0}(y)} \big), E \right) - \| \Delta\| \ge
3g\delta_0 - 2d \ge \eta
$$
with
$$
\eta = 2g\delta_0 \ge 8 d\eu^{4m},
$$
so that $\half \ln \frac{\eta}{4d} \ge 2m$.
Now the Combes--Thomas estimate (cf. \eqref{eq:lem.CT}--\eqref{eq:lem.CT.m}) implies
$$
\max_{z\in\pt \ball_{L_0}(y)}
|G_{\ball_{L_0}(y)}(y,z;E;\om;\th)| \le \eu^{-2m L_0} \le (3L_0)^{-1} \eu^{- \gamma(m, L_0)L_0},
$$
for $L_0$ large enough, hence $\ball_{L_0}(y)$ is \EmNS. The assertion follows.
\qedhere

\section{Separation of local spectra: arbitrary scale}
\label{ssec:sep.spec.Lj}


We will use the following notation: by writing
$\lr{\Lam', \Lam''} \sqsubset \Lam$, we  mean that
$
\Lam', \Lam'' \subset \Lam  \;\; \text{ and } \Lam'\cap\Lam'' = \varnothing.
$
Let
\begin{align}\label{eq:D.L.om.th.1}
D(L, \om, \theta; x, y)
& = \dist\left(\sgomth{L}{x}, \; \sgomth{L}{y} \right) , \\
\label{eq:D.L.om.th.2}
D(L, \om, \theta)
& =
\min_{ \lr{\ball_L(x), \,\ball_L(y)} \sqsubset \ball_{L^{4}}(0) }
\;\;
D(L, \om, \theta; x, y) , \\
\label{eq:D.L.om.th.3}
D(L, \theta)
& =  \inf_{\om\in\Om} \; D(L, \om, \theta) ,
\end{align}
and, for $N\ge 1$,
\begin{align}\label{eq:D.N.L.om.th.1}
D^{(N)}(L, \om, \theta; x, y)
& = \dist\left(\Sigma\big(H^{(N)}_{\ball_L(x)}(\om;\th)\big),
    \; \Sigma\big(H^{(N)}_{\ball_L(y)}(\om;\th) \big) \right),
\\
\label{eq:D.N.L.om.th.2}
D^{(N)}(L, \om, \theta)
& =  \min_{ \lr{\ball_L(x),\ball_L(y)} \sqsubset \ball_{L^{4}}(0) }
\;\;
D^{(N)}(L, \om, \theta; x, y),
\\
\label{eq:D.N.L.om.th.3}
D^{(N)}(L, \theta)
& =  \inf_{\om\in\Om} \; D^{(N)}(L, \om, \theta).
\end{align}

\vskip3mm
\noindent

\subsection{Spectral separation estimates for the local Hamiltonians}
\label{ssec:sep.truncated.fixed.omega}

\bco\label{cor:approx.H.const}
Fix $j\ge 0$. Using the notations of Lemma \ref{lem:loc.const.H} and
\eqref{eq:D.N.L.om.th.1}--\eqref{eq:D.N.L.om.th.3},
with $\Nj = \tN(L_j)$ (cf. \eqref{tN.A.C}),
assume that
for each $\taujl$ one has (cf. \eqref{eq:def.delta.j})
\be\label{eq:cor.approx.H.tN}
D^{(\Nj)}(L_j, \taujl, \th) \ge 5 g \delta_j .
\ee
Then for the non-truncated operators, one has the uniform
lower bound
\be\label{eq:cor.approx.H}
D(L_j,\th) = \inf_{\om\in\Om} \; D(L_j, \om, \th) \ge 4g \delta_j.
\ee
\eco
\proof
By Lemma \ref{lem:loc.const.H}, the condition \eqref{eq:cor.approx.H.tN} implies
$D^{(\Nj)}(L_j, \om, \th) \ge 5 g \delta_j$ for all $\om\in\Om$.
Further, by \eqref{eq:norm.v.vN}, we have
$$
\| H_{\ball_{L^4_j}(u)}(\om;\th) - H_{\ball_{L^4_j}(u)}^{(\Nj)}(\om;\th)\|
\le \half g 2^{-2 b \Nj} a_{\Nj}
= \half  g\delta_j,
$$
so by the min-max principle, the perturbations $|E_i^{\Nj}- E_i|$ of the respective
eigenvalues $E_i\in\Sigma\big(H_{\ball_{L^4_j}(u)}(\om;\th)\big)$ induced by the approximation
of $H_{\ball_{L^4_j}(u)}(\om;\th)$ by $H^{(\Nj)}_{\ball_{L_j^4}(u)}(\om;\th)$,
does not exceed $\half  g\delta_j$.
Consequently, if for a pair of eigenvalues of $H^{(\Nj)}_{\ball_{L_j^4}(u)}(\om;\th)$ we have
$|E_{i'}^{\Nj}- E_{i''}^{\Nj}| \ge 5g\delta_j$ (which follows from
\eqref{eq:cor.approx.H.tN} for all $\om$), then
$$
|E_{i'}- E_{i''}| \ge |E_{i'}^{\Nj}- E_{i''}^{\Nj}| - 2\half g\delta_j\ge 5g\delta_j - g\delta_j
= 4 g \delta_j.
$$
Thus  \eqref{eq:cor.approx.H.tN} implies \eqref{eq:cor.approx.H}.
\qedhere

\vskip1mm

We see that in order to guarantee a lower bound on $D(L_j,\th)$, it suffices
to estimate a finite number of $\th$-probabilities for the approximants of order
$\Nj$ and $\om \in\{ \taujl, 1 \le l\le \cL'_j \}$, where $\cL'_j \le \cL_j$.
This task is performed
in Sect.~\ref{ssec:exclusion.Wegner}.

\subsection{Exclusion of bad $\th$-sets by the Wegner-type estimate}
\label{ssec:exclusion.Wegner}

\begin{lemma}\label{lem:cor.lem.DLom}
Under the assumptions \UPA and \DIV, for any $b > b_* :=(8d + 4\nu A + 4\nu A')/(10 A)$ and
$L_0$ large enough
\be\label{eq:lem.cor.lem.DLom}
 \prTh{ \inf_{\om\in\Om} D(L_j,\th) < 4 g \delta_j} \le   L_j^{- bA }.
\ee
\end{lemma}

\proof
Fix $j\ge 0$ and let $\Nj=\tN(L_j) = O(\ln L_j)$ be given by \eqref{tN.A.C}.
Further, fix a pair of
disjoint cubes $\ball_{L_j}(x)$, $\ball_{L_j}(y) \subset \ball_{L_j^4}(0)$ and consider
the operators $H^{(\tN_j)}_{\ball_{L_j}(x)}$, $H^{(\Nj)}_{\ball_{L_j}(y)}$.
Recall that
all the points of any finite trajectory of the form
$\{T^z\om, \, z\in\ball_{L_j^4}(0)\}$ are separated by the elements
of the partition $\cC_{\Nj}$. Such a separation occurs in particular for
$\big\{T^z\om, z\in \ball_{L_j}(x) \cup \ball_{L_j}(y)\big\}$,
thus conditional on the sigma-algebra $\cB_{\ne \Nj}$ generated by
$\{ \th_{n,k}, \, n\ne \Nj\}$, and with fixed $\om\in\Om$,
the probability distribution of the potential $V_{\Nj}(z;\om,\th)$, generated by the truncated hull
$v_{\Nj}$, gives rise to the sample
of independent random variables (relative to the probability space $\Th$, and not $\Om$ !)
\be\label{eq:cV}
\cV_{\Nj}(\om,\th) := \{ v_{\Nj}(T^z\om;\th), z\in \ball_{L^4}(0) \};
\ee
each of them is uniformly distributed  in its individual interval
$[c_z, c_z +a_{\Nj}]$, with $c_z = c_z(\om,\th)$ determined by the
random (in $\th$) amplitudes $\th_{n,k}$, from generations with $n < \Nj$.
Therefore, conditional on $\cB_{\ne \Nj}$, the independent random variables listed in \eqref{eq:cV}
have individual probability densities, uniformly bounded by $a_{\Nj}^{-1}$.
As a result, conditional on $\fB_{L_j}$, the operators
$H^{(\tN)}_{\ball_{L_j}(x)}$ and $H^{(\Nj)}_{\ball_{L_j}(y)}$ are
independent,
and
for every fixed $\taujl\in\cT_j$, by Theorem \ref{thm:Minami} with $J=1$ (Wegner-type bound),
$$
\bal
&  \prTh{ D^{(\Nj)}(L_j, \taujl, \theta; x, y) \le 5 g\delta_j  }
\\
& \quad
\equiv
\esmth{\prTh{ \dist
\left[ \Sigma\left(H^{(\Nj)}_{\ball_{L_j}(x)}(\taujl;\th)\right),
        \Sigma\left(H^{(\Nj)}_{\ball_{L_j}(y)}(\taujl;\th)\right)
\right] \le 5 g \delta_j \cond \fB_{L_j} } }
\\
& \quad \le \sup_{\lam\in\DR}\;\; \essup \;\;
  \prTh{ \dist
\left[ \Sigma\left(H^{(\Nj)}_{\ball_{L_j}(x)}(\taujl;\th)\right), \lam \right] \le 5 g \delta_j \cond \fB_{L_j} }
\\
& \quad \le 3^{2d} L_j^{2d} a_{\Nj}^{-1} \, 5 \delta_j \cdot g \cdot g^{-1}.
\eal
$$
Since the number of all pairs
$x,y\in\ball_{L_j^4}(0)$ is bounded
by $|\ball_{L_j^4}(0)|^2/2 \le 3^{2d} L_j^{8d}/2$, we obtain (cf. \eqref{eq:D.N.L.om.th.2})
$$
\prTh{ D^{(\Nj)}(L_j, \taujl, \theta) \le 5 g \delta_j }
\le \half 3^{4d} L_j^{8d} a_{\Nj}^{-1} \cdot 5 \delta_j.
$$
Further,
$$
\bal
\prTh{ \min_l D^{(\Nj)}(L_j, \taujl, \theta) < 5g \delta_j}
&\le \cL_j \max_l \prTh{  D^{(\Nj)}(L_j, \taujl, \theta) < 5 g \delta_j}
\\
& \le C(d,\nu) L_j^{8d + 4\nu A + 4\nu A'} a_{\Nj}^{-1} \, \delta_j.
\eal
$$
By Corollary \ref{cor:approx.H.const}, we conclude that
$$
\bal
\prTh{ \inf_{\om\in\Om} D(L_j,\th) < 4g \delta_j} & \le
\prTh{ \min_{l} D^{(\Nj)}(L_j, \taujl, \theta) < 5 g \delta_j}
\\
&\le C_1(d,\nu) L_j^{8d + 4\nu A + 4\nu A'} a_{\Nj}^{-1} \,\delta_j.
\eal
$$
By construction (cf. \eqref{eq:def.delta.j}),
$$
\delta_j =   2^{-2b \Nj} a_{\Nj} \le  C'' (L_j^4)^{- 3 b A} a_{N_j} .
$$
By our assumption, $b = \eps + (8d + 4\nu A + 4\nu A')/(10A)$, $\eps>0$, thus
for large $L_0$,
$$
\bal
\prTh{ \inf_{\om\in\Om} D(L_j,\th) < 4 g \delta_j} & \le
 C_2(d,\nu)  L_j^{8d + 4\nu A + 4\nu A' - 12 b A }
\\
&\le  C_2(d,\nu) L_j^{-\eps}  \cdot  L_j^{- 2b A } \le L_j^{- b A } ,
\eal
$$
which proves \eqref{eq:lem.cor.lem.DLom}.
\qedhere

\vskip1mm

Now define the sets
\be\label{eq:def.Thinf}
\bal
\Th^{(j)}(g) &:= \myset{\th\in\Th:\,  \inf_{\om\in\Om} D\big(L_j,\th\big) \ge 4g \delta_j },
\;\; j\ge 0,
\\
\Thinf(g) &:= \cap_{j\ge -1} \Th^{(j)}(g).
\eal
\ee

\bco\label{cor:mes.Thj}
Under the asumptions \UPA and \DIV, for any $b> b_* := (8d + 4\nu A + 4\nu A')/(10 A)$
and $L_0$ large enough (or for $b>0$ large enough),
\be\label{eq:mu.Th.j}
\forall\, j\ge 0 \qquad \prth{ \Th^{(j)}(g)} \ge 1 -   L_j^{- bA}
\ee
and, therefore, owing to the estimate \eqref{eq:prthp.Thzero},
\be\label{eq:mu.Thinf}
\prth{ \Thinf(g)} \ge 1 - \Const\, \eu^{- c' \ln^{1/2}g} \tto{g\to\infty} 1.
\ee
\eco
\proof
The estimate \eqref{eq:mu.Th.j} follows directly from Lemma \ref{lem:cor.lem.DLom}
and the definition \eqref{eq:def.Thinf} of the set $\Th^{(j)}(g)$. The second assertion
\eqref{eq:mu.Thinf} follows from \eqref{eq:mu.Th.j} by a simple calculation, for
$g$ large enough, since
$\sum_{j\ge 1} L_j^{- bA} < \infty$,
and $L_j(g) = (L_0(g))^{2^j}\to\infty$ as $g\to\infty$.
\qedhere

\subsection{Sparseness of resonant cubes}

Recall (cf. Definition \ref{DefNR}) that, given $E\in\DR$ and a DSO $H_{\ballj(x)}$,
the cube $\ballj(x)$ is
called $E$-resonant if
$$
\dist\Big[ \Sigma\big(H_{\ballj(x)}\big), E \Big] <  g \delta_j.
$$
Taking into account Corollary \ref{cor:mes.Thj},
we come to an important conclusion: for any "good" value of $\th$ and every (not just $\DP$-a.e. !)
$\om\in\Om$, the \ER cubes are sparse:

\bco\label{cor:no.pair.ER}
For
$g$ large enough
and any $(\om,\th)\in\Om\times\Thinf(g)$, for each $j\ge 0$
and any $E\in\DR$, there is no pair of disjoint \ER cubes
$\ball_{L_j}(x)$, $\ball_{L_j}(y)$ $\subset$ $\ball_{L_j^4}(0)$.
\eco

\proof
Assume otherwise; then for some disjoint cubes
in $\ball_{ L_j^4}(0)$
$$
\bal
 \dist
&\left[ \Sigma\left(H^{(\tN_j)}_{\ball_{L_j}(x)}(\om;\th)\right),
        \Sigma\left(H^{(\tN_j)}_{\ball_{L_j}(y)}(\om;\th)\right)
\right]
\\
& \quad \le \left[ \Sigma\left(H^{(\tN_j)}_{\ball_{L_j}(x)}(\om;\th)\right), E \right]
+
\left[E,
        \Sigma\left(H^{(\tN_j)}_{\ball_{L_j}(y)}(\om;\th)\right)
\right]
\\
 & \quad \le  g \delta_j + g \delta_j
<   4 g \delta_j,
\eal
$$
which is impossible for $\th\in\Th^{(j)}(g)$, due to \eqref{eq:def.Thinf}.
\qedhere

\section{Simplified scale induction for deterministic operators}
\label{sec:MSA}

Now we can start collecting the fruits of the tedious analysis of eigenvalue concentration
for the local Hamiltonians $H_{\ball_L(x)}(\om;\th)$, performed in the previous sections.

\subsection{Decay of the Green functions in finite cubes}\label{ssec:decay.GF.balls}

\bde\label{def:ell.q.dominated}
Let $L\ge \ell\ge 0$ be integers and $q\in(0,1)$. Consider a finite set $\Lam\subset\DZ^d$
such that $\Lam\supset\ball_{L+1}(u)$.
A function $f:\,\Lam\to\DR_+$
is called $(\ell,q)$-dominated in $\ball_L(u)$ if for any cube
$\ball_\ell(x)\subset\ball_{L}(u)$ one has
\be\label{eq:def.subh}
|f(x)| \leq q \;\;\mymax{y:\, |y-u| \le\ell +1} |f(y)| .
\ee
\ede

Below we use the notation $\cM(f,\Lam) := \max_{x\in\Lam} |f(x)|$.

The motivation for this definition comes from the following observation.

\ble\label{lem:psi.subh}
Consider a cube $\ball=\ball_{L}(u) \subset\ball_{L+1}(u) \subset \Lam \subseteq \DZ^d$,
$L\ge \ell\ge 0$, $u\in\DZ^d$, and
the operator $H_\Lam = -\Delta_\Lam + gV$ in $\ell^2(\Lam)$ with fixed potential $V$.
Fix $E\in\DR$ and let $\psi\in\ell^2(\Lam)$ be a normalized eigenfunction
of $H_\Lam$ with eigenvalue $E$.
If every cube $\ball_{\ell}(x)\subset\ball$
is \EmNS for some $m \ge 1$, then the function
$
x \mapsto |\psi(x)|
$
(bounded by $1$)
is $(\ell,q)$-dominated
in $\ball$, with $q = \eu^{-\gamma(m,\ell) }$.
\ele

\proof
By the Geometric Resolvent Inequality for the eigenfunctions (cf. \cite{K08}),
$$
|\psi(x)| \le |\ball_{\ell}(x)| \max_{y: |y-x|=\ell} |G_{\ball_\ell(x)}(x,y;E)|
\max_{y: |y-x|=\ell+1} \, |\psi(y)|.
$$
The assumed \EmNS  property of $\ball_\ell(x)\subset\ball$ implies that,
for $\ell\ge 1$, the two maxima figuring in the RHS are bounded, respectively, by
$|\ball_{\ell}(x)|^{-1} \eu^{-\gamma(m,\ell)}$ and by $\|\psi\|_\infty$.
This proves the claim.
\qedhere

\ble[Cf. Lemma 4 in \cite{C12a}]\label{lem:FG.subh}
Consider a cube $\ball=\ball_{L_{k+1}}$, $k\ge 0$, and $H_{\ball}$ with fixed potential $V$. Pick
$x_0,y_0\in\ball$ with $|x_0 - y_0| > L_{k}$, and fix $E\in\DR$.
Suppose that $\ball$ is $E$-NR and every cube $\ball_{L_{k}}(x)\subset\ball$
is \EmNS for some $m \ge 1$. Then the function
$$
f_{y_0}: x \mapsto |G_{\ball}(x,y_0;E)|
$$
is $(L_{k},q)$-dominated
in $\ball$, with $q = \eu^{-\gamma(m,L_{k}) }$,
and bounded by $\eu^{L_k^\beta}$.
\ele
The proof is similar to that of Lemma \ref{lem:psi.subh} and will be omitted; the upper bound on $f_{y_0}$
follows, of course, from the \ENR property.

\ble[Cf. Lemma 2 in \cite{C12a}]\label{lem:subh.1} Suppose that a function $f:\Lam\to\DR_+$, with
$\DZ^d\supset\Lam\supset\ball_{L+1}(x)$, is $(\ell,q)$-dominated in $\ball_L(x)$.
Then
$$
 |f(x)| \le q^{ \left\lfloor \frac{L+1}{\ell+1} \right\rfloor } \cM(f,\Lam)
 \le q^{ \frac{L - \ell}{\ell+1} } \cM(f,\Lam) .
$$
\ele

We omit the proof, the details of which can be found in Refs. \cite{C12a} and \cite{CS13}.
\vskip1mm

\noindent
$\blacklozenge$ We stress that the values $L=0$ and $\ell=0$ are indeed admissible.

\bde
A cube $\balljone(u)$, $j\ge 0$, is called \mbad, if for some $E\in\DR$,
it contains at least
two disjoint \EmS cubes of radius $L_j$. Otherwise, it is called \mgood.
\ede

The following statement is a (simpler)
variant of Lemma 4.2 in \cite{DK89}; similar results have been
used in numerous papers using the Multi-Scale Analysis; cf. e.g.,
Lemma 4.4 in \cite{Dr87},
or Theorem 10.14 and a stronger Theorem 10.20 in the review \cite{K08} by Kirsch, or Theorems 2.4.1, 2.4.3
and Lemma 2.4.4 in \cite{CS13}, or Lemma 5 in \cite{C14a}.
For these reasons, and for brevity, we omit the proof.

\ble\label{lem:NT.NR.is.NS}
For $m\ge 1$ and $L_0$ large enough,
if a cube $\ball_{L_{j+1}}(u)$, $j\ge 0$, is \mgood  and
\ENR for some $E\in\DR$, then it is \EmNS.
\ele

Introduce the following property which will be proved by scale induction:

\vskip1mm
\sparsej{}:
\textit{
For all $\th\in\Thinf(g)$, $\om\in\Om$, $E\in\DR$ and $u\in\DZ^d$,
the cube $\ball_{L^4_j}(u)$ contains no pair of disjoint
$(E,m;\om;\th)${\rm-S} cubes of radius $L_j$.
.
}
\vskip2mm

Recall that we set $L_{-1} = 0$; it is convenient to formulate in a special way
the property \sparseL{L_{-1}} $\equiv$ \sparsez:

\vskip2mm
\sparsez:
\textit{
For all $\th\in\Thinf(g)$, $\om\in\Om$, $E\in\DR$ and $u\in\DZ^d$,
there is at most one  point $x \in \ball_{L^4_0}(u)$ such that the single-site cube
$\ball_{0}(x)$ is $(E,m;\om;\th)${\rm-S}.
}
\vskip2mm

The property \sparsej could be formulated in an equivalent way, where only the cubes $\ball_{L_j}(0)$
(centered at the origin)
are considered, since $H_{\ball_{L_j}(u)}(\om;\th)$ $=$ $H_{\ball_{L_j}(0)}(T^u\om;\th)$.

\vskip1mm

Anticipating the discussion in Sect.~\ref{sec:Uniform},
we can say that the "exceptional" sites mentioned
in \sparsez will be the centers of localization of  unimodal eigenfunctions, with eigenvalues $E$
``close'' to $gV(T^x\om;\th)$: $E = gV(T^x\om;\th) + O(\|\Delta\|)$.

For $g$ large enough (i.e.,  with $m = m(g) \gg 1$), the property
\sparseL{L_{-1}} $\equiv$ \sparseL{0} follows directly from Lemma \ref{lem:sep.L0}, since
$\Thinf(g)\subset\Th^{(-1)}(g)$.

\btm\label{thm:dichotomy.1} Assume that
\sparsej holds for some $j\ge 0$.  Then \sparsejone also holds true.
Consequently, \sparsepar{0} implies \sparsej for all $j\ge 0$.
\etm

\proof

Fix any $\th\in\Thinf$, any $u\in\DZ^d$ and any $E\in\DR$. Consider the cube
$\ball_{L^4_{j+1}}(u)$. By definition of the set $\Th^{(j+1)}(g)\supset\Thinf(g)$
and Corollary \ref{cor:no.pair.ER},
there is at most one $E$-R cube $\ball_{L_{j+1}}(v)\subset\ball_{L^4_{j+1}}(u)$.
Let us show by contraposition that there can be no pair of disjoint $(E,m)$-S cubes
$\ball_{L_{j+1}}(x)$, $\ball_{L_{j+1}}(y)$ $\subset$ $\ball_{L^4_{j+1}}(u)$.

Assume otherwise; then one of these cubes -- w.l.o.g., let it be $\ball_{L_{j+1}}(x)$ -- must
be $E$-NR. Then by Lemma \ref{lem:NT.NR.is.NS}, the  cube $\ball_{L_{j+1}}(x)$
must contain two disjoint $(E,m)$-S
cubes of radius $L_j$, which contradicts the hypothesis \sparsej.
\qedhere

\vskip2mm
The property \sparsej, established at all scales $L_j$, $j\ge -1$, uniformly in $\om\in\Om$,
is a stronger -- deterministic -- analog of the well-known probabilistic
"double-singularity" bound for the pairs of \EmS cubes, which represents the final
result of the variable-energy MSA for random operators (cf., e.g., \cite{DK89}).

\subsection{From the MSA to strong dynamical localization}

It would not be difficult now to infer from the results of the deterministic Multi-Scale Analysis,
carried out in the previous subsection, strong dynamical localization for the operators
$H(\om;\th)$. Recall that for random Schr\"{o}dinger operators the derivation of dynamical localization
from the MSA bounds was obtained by Germinet and De Bi\`{e}vre \cite{GDB98}, by Damanik and
Stollmann \cite{DS01}
(in a stronger form) and by Germinet and Klein \cite{GK01} (in yet a stronger
form, and for a larger class of random operators).

However, it will be even easier to infer in Sect.~\ref{sec:DL.uniform}
pointwise and uniform in $\om\in\Om$ dynamical localization
from the uniform (and not just semi-uniform, as in the theory of random Anderson Hamiltonians)
decay of all eigenfunctions, proven in Sect.~\ref{sec:Uniform}.

\section{Uniform localization and unimodal eigenstates}
\label{sec:Uniform}

\bde\label{def:loc.center}
Let $\psi\in\ell^2(\DZ^d)$. A point $x\in\DZ^d$ is called a localization center
for $\psi$ iff $|\psi(x)| = \|\psi\|_\infty$.
\ede

\bde
A normalized eigenfunction $\psi$ of a DSO $H$ is called uniformly $m$-localized if
\begin{enumerate}[\rm(a)]
  \item $\psi$ has a localization center $\hx$ such that $|\psi(\hx)|^2 > \half$;
  \item $\forall\, y\in\DZ^d\setminus \{\hx\}$, one has  $|\psi(y)| \le \eu^{-m |x -y|}$.
\end{enumerate}
When the value $m$ is irrelevant, we will simply say that $\psi$ is uniformly localized.
\ede

Sometimes we will refer to (a) as the unimodality property of $\psi$.

Note that every normalized eigenfunction in $\DZ^d$
admits a non-empty but finite set of its localization centers; it will be denoted by $\hX(\psi)$.
As shows assertion (A) of Lemma \ref{lem:unimod} below, with no loss of generality,
we can restrict our analysis to the situation where the localization center is unique, so
we will write $\hx(\psi)$.

\ble\label{lem:unimod}
\begin{enumerate}[\rm(A)]
  \item Any uniformly localized eigenfunction $\psi$ of a DSO
  $H$ has a unique localization center.

  \item Let $\{\psi_i, i\in\cI\}$, $\cI\subset\DN$, be an orthonormal family of uniformly localized
  eigenfunctions of a given DSO $H$. Then for any $x\in\DZ^d$, there is at most one eigenfunction $\psi_i$
  with localization center $x$.
\end{enumerate}

\ele

\proof
(A)
By Definition \ref{def:loc.center}, $|\psi(x)|$ takes the constant
value $\|\psi\|_\infty$ at all its localization centers $x\in\hX(\psi)\ne \varnothing$.
It follows from the condition (a) of the uniform localization that
$|\psi(x)|^2  > \half$ for $x\in\hX(\psi)$.
By normalization,
$$
\bal
1 &= \sum_{y\in\hX(\psi)} |\psi(y)|^2 + \sum_{y\not\in\hX(\psi)} |\psi(y)|^2
\ge |\hX(\psi)| \cdot |\psi(\hx)|^2
> \half |\hX(\psi)|,
\eal
$$
yielding $|\hX(\psi)| < 2$.

\noindent
(B) Assume otherwise, and let $\phi, \psi$ be orthogonal, normalized,
uniformly localized eigenfunctions of $H$ with localization center $x$, and let
$\chi = \one_{\DZ^d\setminus\{x\}}$.
Then we have $\| \chi \phi\|^2_2$, $\| \chi \psi\|^2_2< 1/2$, thus by Cauchy--Schwarz
inequality,
$$
\bal
| (\phi,\psi)| &= \Big| \phi(x) \psi(x) + \sum_{y\ne x} \phi(y) \psi(y) \Big|
\ge | \phi(x)| \cdot | \psi(x)| - \Big| \sum_{y\ne x} \phi(y) \psi(y) \Big|
\\
& > \frac{1}{\sqrt{2}} \cdot \frac{1}{\sqrt{2}} -  \| \chi \phi\|_2 \, \| \chi \psi\|_2
> \half - \half = 0,
\eal
$$
so that $\phi$ and $\psi$ are not orthogonal; this contradiction proves the claim.
\qedhere

\vskip1mm

In view of Lemma \ref{lem:unimod}, given an eigenbasis $\{ \psi_i, i\in\cI\}$
of uniformly localized eigenfunctions
of a DSO $H$, we can associate with each localization center $\hx$
of some uniformly localized eigenfunction $\psi_i$ a unique eigenvalue $\hlam = \hlam(\hx)$ -- the one of
the eigenfunction $\psi_i$. In Sect.~\ref{sec:Minami}, we will show that,
for typical $\th$ and all $\om\in\Om$, the mapping $\hx \mapsto \hlam(\hx)$ is actually
a bijection, since the spectrum of $H(\om;\th)$ for such $\th$ is simple (and pure point).

To prove that every $x\in\DZ^d$ is a localization center for some eigenfunction of $H(\om;\th)$
(cf. Theorem
\ref{thm:complete.unimodal.EF}), we will need the following simple auxiliary result, valid for
any DSO, regardless of the form of its potential.

\ble\label{lem:loc.center.S}
Let $\psi$ be a normalized eigenfunction of a DSO $H$, and let $\hx$ be
any of its localization centers.
Then for any $L\in\DN$, the cube $\ball_L(\hx)$ is {\rm$(\hlam(\hx),m)$-S}.
\ele
\proof
Fix an eigenfunction $\psi$ with localization center $\hx$ and assume otherwise.
Since $\gamma(m,L)>0$ and $q:=\eu^{-\gamma(m,L)}<1$, Lemma \ref{lem:psi.subh} implies
$$
\|\psi\|_\infty =
|\psi(\hx)| \le \eu^{-\gamma(m,L)L} \max_{y\in\pt^+ \ball_L(\hx)} |\psi(y)|
   \le q \, \|\psi\|_\infty ,
$$
thus $\|\psi\|_\infty = 0$, which is impossible, since $\|\psi\|_2 >0$.
\qedhere

\ble\label{lem:EF.unif.loc}
Consider a DSO $H$ and assume that \sparsej holds true for all $j\ge -1$, and $L_0 \ge 11$.
If, in addition, $m>0$ is large enough, so that
\be\label{eq:sum.half}
\sum_{ r\ge 1} (3  r)^{d} \eu^{-  m  r} \le \half ,
\ee
then every normalized eigenfunction $\psi$ of $H$,
with localization center $\hx$, is uniformly $m$-localized at $\hx$.
Furthermore, any polynomially bounded solution $\Psi$ to the equation $H(\om;\th)\Psi = E\Psi$
decays exponentially fast, thus $\|\Psi\|_2 < \infty$; consequently, $H(\om;\th)$ has pure point spectrum.
\ele
\proof

\noindent
$\bullet$ \textbf{Step 1.}
Fix an eigenfunction $\psi$ with $\|\psi\|_2=1$, $\hx\in\hX(\psi)$,
$H\psi = \hlam \psi$,
and assume first that $R:=|y - \hx|\in[1,L_1]$, $L_1 = L_0^2<L_0^4$.
By Lemma \ref{lem:loc.center.S}, the  cube $\ball_0(\hx) = \{\hx\}$
is $(\hlam,m)$-S.
Therefore,
by \sparseL{0},
for all $u$ with $|\hx -u|\in[1,L_1]$,
the single-site cubes $\ball_0(u) = \{u\}$
are $(\hlam,m)$-NS. Fix any $y$ with $1 \le R := |\hx -y| \le L_1$ and set $r := R-1$.
Each single-site cube $\ball_{0}(u)\subset\ball_{r}(y)$ is $(\hlam,m)$-NS, so by Lemma \ref{lem:subh.1}
(where one has to set $L=r$, $\ell=0$), combined with Lemma \ref{lem:psi.subh}, we have
$$
|\psi(y)| \le \eu^{-\gamma(m,0) \left\lfloor \frac{r+1}{0+1} \right\rfloor} \|\psi\|_\infty
\le \eu^{-\gamma(m,0) (r+1)} \le \eu^{-2 m |y - \hx|}.
$$
Using \eqref{eq:sum.half} and the crude estimate $\card\{u:\, |u|=r\}\le (2r+1)^d \le (3r)^d$, we obtain
\be\label{eq:psi.A0}
\bal
\sum_{y\in\ball_{L_1}(\hx)\setminus \{ \hx \}} |\psi(y)|^2
&\le \sum_{r=1}^{L_1} (3 r)^d \eu^{- 4 m r}.
\eal
\ee

\vskip1mm
\noindent
$\bullet$ \textbf{Step 2.}
Now let $R:=|y-\hx|>L_1$.
The complement of $\ball_{L_1}(\hx)$ is covered by the disjoint annuli:
$$
\CZ \setminus \ball_{L_1}(\hx) = \bigcup_{j \ge 2} \rA_j, \quad
\rA_j := \ball_{L_{j}}(\hx) \setminus \ball_{L_{j-1}}(\hx).
$$
Fix $j\ge 2$ and $y\in\rA_j$, so $R > L_{j-1}$.
Since $\ball_{L_{j-2}}(\hx)$ is $(\hlam,m)$-S,
every cube
$$
\ball_{L_{j-2}}(u)\subset \ball_{R-L_{j-2}}(y)
   \subset \ball_{L_{j-2}^4}(\hx) \setminus \ball_{L_{j-1}}(\hx) ,
$$
being disjoint from $\ball_{L_{j-2}}(\hx)$,
must be $(\hlam,m)$-NS, owing to \sparsepar{j-2}.
By Lemma \ref{lem:psi.subh} and Lemma \ref{lem:subh.1},
with $\|\psi\|_\infty \le 1$,
$$
\bal
|\psi(y)| &\le \eu^{-m(1+L_{j-2}^{-1/8}) L_{j-2} \cdot \frac{(R-L_{j-2}) -2L_{j-2}}{L_{j-2}+1} }
   \|\psi\|_\infty
\\
&
  \le \eu^{-m R \, \left(1+L_{j-2}^{-1/8} \right) \frac{ 1 - 3L^{-1}_{j-2} }{1 + L_{j-2}^{-1}}  }
\le \eu^{-m R \, \left(1+L_{j-1}^{-1/8} \right) \left(1 - 4L^{-1}_{j-1} \right)  }
\\
&
< \eu^{- m R } ,
\eal
$$
provided that $11 \le L_0 \le L_{j-1}$, as shows an elementary numerical
calculation\footnote{It suffices that $L_0^{7/8} \ge 8$, and actually $11^{7/8}>8$.}.
Since $|\rA_j[\le ( 2L_{j} + 1)^d  \le  (3 L_{j})^{d}$, we obtain,
with $R > L_j$,
\be\label{eq:psi.Aj}
\bal
\sum_{y\in\rA_j } |\psi(y)|^2
&\le   (3 L_{j})^{d} \left(\eu^{- m L_j}\right)^2.
\eal
\ee
Collecting  \eqref{eq:psi.A0}, \eqref{eq:psi.Aj} and \eqref{eq:sum.half}, we conclude
that
$$
\bal
\sum_{y\ne \hx } |\psi(y)|^2
&\le \sum_{r=1}^{L_1} (3r)^d \eu^{-4 m r}
   + \sum_{j\ge 2} (3 L_{j})^{d} \eu^{- 2 m L_j}
\\
&\le  \eu^{-m}\, \sum_{r=1}^{\infty} (3 r)^{d} \eu^{- m r}
\le \half \,\eu^{-m} < \half.
\eal
$$
Therefore, $|\psi(\hx)|^2>1/2$, so $\psi$ is uniformly $m$-localized at $\hx$.

For the proof of the second assertion, it suffices to repeat Step 2, but replace the uniform bound
$\|\psi\|_\infty \le 1$ by $|\psi(z)|\le C (|z|+1)^a$, $z\in\DZ^d$,
and also replace  $\hx$ by any point $\hy$ where $\Psi(\hy)\ne 0$.
This still gives an exponential upper bound on $|\Psi(y)|$ in
$\ball_{L_j}(\hy)\setminus\ball_{L_{j-1}}(\hy)$, for $j$ large enough.\footnote{In both cases (normalized eigenfunctions and polynomially bounded
generalised eigenfunctions), the argument we use is well-known and goes back to \cite{DK89,FMSS85}.}
It is well-known (cf. e.g., \cite{K08}) that for spectrally-a.e. $E\in\Sigma(H)$, a DSO $H$
has a polynomially bounded generalized eigenfunction with eigenvalue $E$. Therefore,
$H(\om;\th)$ has pure point spectrum for all $(\om,\th)\in\Om\times \Thinf(g)$.
\qedhere

\vskip3mm

The following statement marks the end of the proof of Theorem \ref{thm:Main}.

\btm\label{thm:complete.unimodal.EF}
For all sufficiently large $m \ge 1$ and $g\ge g_*(m)$  large enough,
so that in particular \eqref{eq:sum.half} holds true, for
any $(\th,\om)\in\Thinf(g)\times \Om$, the operator $H(\om;\th)$
has an eigenbasis of uniformly $m$-localized eigenfunctions $\psi_x$, uniquely
labeled by their respective localization centers:
$$
\forall\, x\in\CZ\qquad
\hX(\psi_x) = \{x\}, \;\; |\psi_x(x)|^2 > 1/2 .
$$
For any $x\in\CZ$ there is exactly one eigenfunction of $H(\om;\th)$ localized at $x$.
\etm

\proof
By Lemma \ref{lem:EF.unif.loc}, for all $g$ large enough, $H(\om;\th)$ has an eigenbasis of uniformly $m$-localized eigenfunctions $\psi_k$, $k=1, 2, \ldots$.
Therefore, by Lemma \ref{lem:unimod}, each $\psi_k$ admits a unique
localization center $\hx(\psi_k)$. It remains to show that each point $x\in\DZ^d$ is the localization center
for exactly one eigenfunction.

Pick any $x\in\DZ^d$;  then we have
by the Parseval identity:
$$
\bal
1 &
= \sum_k |\psi_k(x)|^2
= \sum_{k:\, x\in\hX(\psi_k)} |\psi_k(x)|^2 + \sum_{k:\, x\not\in\hX(\psi_k)} |\psi_k(x)|^2
=: S_1 + S_2.
\eal
$$

By assertion (B) of Lemma \ref{lem:unimod},
distinct uniformly $m$-localized  eigenfunctions have distinct localization centers,
thus
$$
\bal
S_2 &= \sum_{k:\, x\not\in\hX(\psi_k)} |\psi_k(x)|^2
\le \sum_{r=1}^{\infty} \;\;\sum_{k:\, |x - \hx(\psi_k)| = r} \eu^{-2m r}
\le \sum_{r=1}^{\infty} (3 r)^{d} \eu^{-2m r}
\\
&
\le \eu^{-m } \cdot \frac{1}{2} < 1 \quad \Longrightarrow \quad S_1 >0.
\eal
$$
Hence $1 \ge |\{k:\, x\in\hX(\psi_k)\}| >0$ for any $x\in\DZ^d$, so
$|\{k:\, x\in\hX(\psi_k)\}|=1$, and there exists a bijection
between the elements $\psi_k$ of the eigenbasis of uniformly $m$-localized, unimodal
eigenfunctions and the lattice $\DZ^d$.
\qedhere
\section{Uniform dynamical localization}
\label{sec:DL.uniform}

Below we use the standard Dirac's "bra-ket" notation $\langle \phi | H | \psi \rangle$
for the scalar product of $\phi$ and $H\psi$ in the Hilbert space $\ell^2(\DZ^d)$.
\btm\label{thm:DL.uniform}
For all $g>0$ large enough, all $\th\in\Thinf(g)$, for any $\om\in\Om$ and all $x,y\in\DZ^d$,
for any continuous function $\phi:\DR\to\DC$ with $\|\phi\|_\infty\le 1$,
$$
|\langle \one_x | \phi(H(\om;\th)) | \one_y \rangle| \le \Const(d)\, |x-y|^d \eu^{-m |x-y |}.
$$
\etm

\proof
By functional calculus,
we have the following identity, assuming that the series in the RHS of
\eqref{eq:proof.DL.1} converges absolutely:
\be\label{eq:proof.DL.1}
\langle \one_x | \phi(H) | \one_y \rangle
= \sum_{z\in \DZ^d} \langle \one_x  | \psi_z \rangle \, \phi(\lam_z) \, \langle \psi_z| \one_y \rangle,
\ee
so
it suffices to prove convergence of the series
$$
\bal
\|\phi\|_\infty \sum_{z\in \DZ^d} | \langle \one_x  | \psi_z \rangle \langle \psi_z| \one_y \rangle|
\le  \sum_{z\in \DZ^d} | \langle \one_x  | \psi_z \rangle \langle \psi_z| \one_y \rangle|.
\eal
$$
By Theorem \ref{thm:complete.unimodal.EF}, we have
$|\psi_z(x)| \le \eu^{-m|z-x|}$ and $|\psi_z(y)| \le \eu^{-m|z-y|}$,
with the decay exponent $m\ge m_*(g)\to+\infty$ as $g\to+\infty$,
so that
$$
\bal
 \sum_{z\in \DZ^d} | \langle \one_x  | \psi_z \rangle \langle \psi_z| \one_y \rangle| \le
\sum_{z\in \DZ^d}  \eu^{-m |x-z | - m |z-y |}.
\eal
$$
Let $R = |x-y|$. For any  $z\not\in\ball_{2R}(x)$, setting $n = |z-x|\ge 2R+1$,
we have
$$
|z - x| + |z - y| \ge n + \dist(z, \ball_R(x)) \ge 2n -R,
$$
since $y\in \ball_R(x) \subset \ball_{2R}(x) \not\ni z$. Furthermore,
$$
\forall\, n > R\quad \card \{z\in\DZ^d:\, |z - x| = n \} \le C(d) n^{d-1},
$$
thus
$$
 \sum_{z \not\in \ball_{2R}(x)}  \eu^{-m |x-z| -m|z-y|}
 \le \sum_{n > 2R} C(d) n^{d-1} \eu^{-m(2n-R)} \le C'(d) R^{d} \eu^{-2mR}.
$$
For $z\in\ball_{2R}(x)$ (indeed, for any $z\in\DZ^d$) one can use a simpler bound:
by the triangle inequality,
$
|z - x| + |z - y| \ge |x - y|  =R.
$
Therefore,
$$
\sum_{z \in \ball_{2R}(x)} \eu^{-m|x-z| -m|z-y|} \le \eu^{-mR} |\ball_{2R}(x)|
\le  C'' R^d \eu^{-mR}.
$$
Finally,
$$
|\langle \one_x | \phi(H) | \one_y \rangle| \le \Const(d)\, |x-y|^d \eu^{-m|x-y|}.
$$
\qedhere

The standard form of dynamical localization is obtained with the functions
$\phi = \phi_t: \lam \mapsto \eu^{-\ii \lam t}$, $t\in\DR$.

\section{Minami-type estimates. Simplicity of spectra}
\label{sec:Minami}

\subsection{Spectral spacings in large cubes}
\label{ssec:proof.Minami.Wegner}

Recall the generalized Minami estimate \cite{Mi96} proven in Refs \cite{BHS07,GV07}.

\btm[Cf. \cite{BHS07,GV07}]\label{thm:Minami.rand}
Let $H_{\ball_L(u)}(\th)$ be a random DSO relative to some probability space
$(\widetilde{\Th},\widetilde{\fB},\widetilde{\DP})$,
with IID random potential $V(x;\th)$. Assume that the common probability distribution of the random
variables $V(x;\th)$
has a bounded density $\rho$. Then for any finite interval $I\subset\DR$ one has
\be
\widetilde{\DP} \left\{  \Tr \Pi_I( H_{\ball_L(u)}(\th)) \ge J \right\}
\le  \frac{(\pi \|\rho\|_\infty)^J}{ J! } |I|^J .
\ee
\etm

Theorem \ref{thm:Minami.prob.Om.Th.1} is actually an adaptation of Theorem 5.2
from \cite{C11c} to the "haarsh" deterministic potentials.

\proof[Proof of Theorem \ref{thm:Minami.prob.Om.Th.1}]
We prove first \eqref{eq:thm.Minami.prob.Om.Th}: for any fixed $\om\in\Om$,
\be\label{eq:thm.Minami.prob.Om.Th.again}
\prth{ \Tr \Pi_I( H_{\ball_L(0)}(\om;\th)) \ge J} \le  C_J\, L^{J B \ln L} |I|^J .
\ee
To this end, consider the sigma-algebra $\fB_{L_j}$ figuring in \LVB.
Conditional on $\fF\times\fB_{L_j}$ (hence, with fixed $\om$),
the values of the potential $v(T^x\om;\th)$ with $x\in\ball_{L_j}(0)$ become
independent and admit probability densities $\rho_{x,L_j}$ with
$\|\rho_{x,L_j}\|_\infty \le C'' L_j^{B \ln L_j}$. Now the
assertion follows from Theorem \ref{thm:Minami.rand}
applied to the operators $H_{\ball_{L_j}(0)}(\om;\th)$, with fixed $\om$ and subject to
the conditional measure $\prth{\cdot\,|\, \fB_{L_j}}$ with
respect to $\th\in\Th$:
$$\bal
&\prth{ \Tr \Pi_I( H_{\ball_{L_j}(0)}(\om;\th)) \ge J}
\\
& \quad
=
\esmth{ \prth{ \Tr \Pi_I( H_{\ball_{L_j}(0)}(\om;\th)) \ge J \,|\, \fF\times\fB_N} }
\\
&\quad \le {\frac{1}{ J! } \big(\pi C'' L_j^{B \ln L_j}} \big)^J \, |I|^J ,
\eal
$$
which proves the first assertion \eqref{eq:thm.Minami.prob.Om.Th}.

To prove \eqref{eq:thm.Minami.prob.Om.Th.2}, one can repeat the above argument, but replace
$\prth{\cdot}$ by the product measure $\promth{\cdot}$,
and apply the standard identity
$\promth{ \, \cdot \,}$
$=$
$\esmomth{  \promth{ \, \cdot \, \,|\, \fF\times\fB_L } }.
$
Conditioning on $\fF$ is equivalent
to fixing $\om\in\Om$, so we can make use of the first assertion, valid for each $\om\in\Om$.
\qedhere
\vskip1mm

For the proof of Theorem \ref{thm:simple.spectrum}, we also need a bound deterministic
in $\om\in\Om$; it will be proved only for sufficiently small intervals $I_j$.
%
The next statement establishes a lower bound on the spacings $\Sepb{H_{\ball_{L_j}(0)}(\om;\th)}$
(cf. Sect.~\ref{ssec:sep.spec.L0})
uniform  in $\om\in\Om$.

\btm
\label{thm:Minami.determ.Th}
Let the parameter $b>0$ in the definition of the sequence $\{a_n\}_{n\ge 0}$
(cf. \eqref{eq:def.an.b})
be large enough, so that $4A^2b - 2(B + 4A + 4A') > 1$.
Then there exists $j_\circ = j_\circ(g)$ such that for all $j\ge j_\circ$, one has
\be\label{eq:thm.Minami.determ.Th}
\prth{ \inf_{\om\in\Om} \Sepb{( H_{\ball_{L_j}(0)}(\om;\th) } \le  g\delta_j }
\le   L_j^{-  \ln L_j}.
\ee
\etm

\proof
Fix $j\ge 0$, let
$\tN_j = \tN(L_j,A,C)$ (cf. \eqref{tN.A.C}), and
$\ball_j = \ball_{L_j}(0)$.

Consider first
$H^{(\tN_j)}(\om;\th)$ with the truncated potential $V_{\tN_j}(x;\om;\th) = v_{\tN_j}(T^x\om;\th)$.
Next, cover $\Om$
by the sets $\rP_{j,l} \ni \taujl$, $1 \le l \le \cL'(j) \le \cL_j=\Const  L_j^{4A+4A'}$,
introduced in Sect.~\ref{ssec:sep.truncated.fixed.omega} (cf. Lemma \ref{lem:loc.const.H}).
Let $I$ be an interval of length $4\delta_j$.
By \eqref{eq:thm.Minami.prob.Om.Th.again} with  $\om = \taujl$, we have:
$$
\bal
\prth{\min_{l} \;  \Tr \Pi_I( H^{(\tN_j)}_{\ball_j}(\taujl;\th)) \ge 2}
&\le \cL_j C'_J L^{2 B \ln L_j} \delta_j^2
\le C_J {L_j}^{2 B' \ln L_j} \delta_j^2 .
\eal
$$
$H^{(\tN_j)}_{\ball_j}(\om;\th)$ is constant in $\om$ on each set $\rP_l$, and $\cup_l\rP_{j,l} =\Om$,
so the above bound implies that
\be\label{eq:prth.trace.tN}
\prth{\sup_{\om\in\Om}\;  \Tr \Pi_{I}( H^{(\tN_j)}_{\ball_j}(\om;\th)) \ge 2}
\le  C_J L_j^{ 2 B' \ln L_j} \delta_j^2.
\ee

Further, $\| H(\om;\th)\| \le C g$, thus
$\Sigma(H(\om;\th)\subset \cI(g) = [-gE^*, gE^*]$, for some $E^*\in(0,+\infty)$,
so for our purposes, it suffices to analyse only the sub-intervals  of $\cI(g)$.

Next, cover the interval $\cI(g)$ redundantly by
$\cK_j := \left\lfloor \frac{2gE^*}{2\delta_j}\right\rfloor+1 \le C g \delta_j^{-1}$ sub-intervals
of length $4\delta_j$,
$$
I_{j,k} := \big[ -gE^* + 2k \delta_j, \;-gE^* + (2k+4) \delta_j \,\big], \;\; k=0, 1, ...,  \cK_j-1 .
$$
Then every subinterval of length $2\delta_j$ of $\cI(g)$ is covered by at least one
of these intervals $I_{j,k}$. Thus
the $\prthp$-probability that at least one interval of length $2\delta_j$ in $\DR$ contains
for some $\om\in\Om$ at least two eigenvalues of $H^{(\tN_j)}_{\ball_j}(\om;\th)$,
is bounded by (cf. the definition of $\delta_j$ in \eqref{eq:def.delta.j})
$$
\cK_j C_J L_j^{ 2 B' \ln L_j} \delta_j^2 \le \frac{C g \delta_j^2}{\delta_j} L_j^{ 2 B' \ln L_j}
\le C g  L_j^{ - (4A^2 b - 2B') \ln L_j} \le g L_j^{ - \ln L_j}
$$
provided that $4A^2b - 2(B + 4A + 4A') > 1$,
and $j\ge j_\circ(g)$, for $j_\circ(g)$ large enough.
Finally, by the min-max principle, the eigenvalue perturbations,
$\big|E_i - E^{(\tN_j)}_i \big|$,  are bounded
by $\|H_{\ball_j} - H^{(\tN_j)}_{\ball_j}\| \le g\|v - v_{\tN_j}\|_\infty \le \half g\delta_j$,
thus, by the triangle inequality,
$$
\bal
\SepB{ H_{\ball_j} } &\ge
\SepB{ H^{(\tN_j)}_{\ball_j} } - 2 g\|v - v_{\tN_j}\|_\infty
\ge 2g\delta_j - g\delta_j = g\delta_j.
\eal
$$
So, we have proved the following implication: denoting $I_s = [s, s + 4\delta_j] $,
$$
\sup_{\om\in\Om}\;\; \sup_{I_s \subset\cI(g)} \;
   \Tr \Pi_{I_s}( H^{(\tN_j)}_{\ball_j}(\om;\th)) < 2
\;\;\Rightarrow \;\;
\inf_{\om\in\Om}\; \SepB{ H_{\ball_{L_j}(0)}(\om;\th) } \ge g\delta_j \, .
$$
Now the assertion follows from the estimate \eqref{eq:prth.trace.tN}.
\qedhere

\bre
The requirement $j\ge j_\circ(g)$ in Theorem \ref{thm:Minami.determ.Th}
and in Corollary \ref{cor:Minami}
becomes unnecessary for even larger $b>0$, i.e.,
for $b$ large enough, one can set $j_\circ=0$.
\ere

Introduce the sets (here the subscript "M" stands for "Minami")
\be\label{eq:def.ThinfMj}
\bal
\Th_{\rM}^{(j)}(g) &:=
\myset{ \inf_{\om\in\Om} \SepB{ ( H_{\ball_{L_j}(0)}(\om;\th) } \ge  \delta_j } \cap\Thinf(g),
\\
\ThinfM(g) &:= \bigcap_{j\ge j_\circ(g)} \Th_{\rM}^{(j)}(g).
\eal
\ee

\bco\label{cor:Minami}
Under the assumptions
of Theorem \ref{thm:Minami.determ.Th}, for any $j\ge j_\circ(g)$,
\be\label{eq:cor.Minami.determ.Th.j}
\prth{ \Th_{\rM}^{(j)}(g) } \ge 1 -  L_j^{- \ln L_j}
\ee
and therefore, if $g$ is large enough, owing to \eqref{eq:mu.Thinf},
one has
\be\label{eq:cor.Minami.determ.Th.inf}
\prth{ \ThinfM(g)}  \tto{g\to+\infty} 1.
\ee
\eco
\proof
The first assertion follows directly from Theorem \ref{thm:Minami.determ.Th}
and the definition \eqref{eq:def.ThinfMj} of $ \Th_{\rM}^{(j)}(g)$.
With $g$ large small enough (hence, $L_j=L_j(g)$ large enough), the bound \eqref{eq:cor.Minami.determ.Th.inf} follows from \eqref{eq:cor.Minami.determ.Th.j}
by an elementary calculation, since
$\sum_j L_j^{-\ln L_j}<\infty$.
\qedhere

\subsection{The Klein--Molchanov argument. Proof of Theorem \ref{thm:simple.spectrum}}

\ble[Cf. Lemma 1 in \cite{KM06}]\label{lem:KM.Lemma.1}
Let $E$ be an eigenvalue of the discrete Schr\"{o}\-din\-ger operator
$H = -\Delta + V$ in $\ell^{2}(\DZ^d)$ with two linearly independent
eigenfunctions $\ffi_1, \ffi_2\in\ell^{2}(\DZ^d)$ such that for some $\beta>d/2$
and some function $f:\DR_+\to\DR_+$, satisfying $f(r) \le C'r^{-\beta}$, $\beta > d/2$,
one has
$$
\forall\, x\in\DZ^d\qquad |\ffi_j(x)| \le f(|x|)
\;\ j=1,2.
$$
Then there exists $C\in(0,+\infty)$ such that, setting
$\eps_L := Cf(L) L^{d/2}>0$ and
$I_L$ $ = [E - \eps_L, E + \eps_L]$ one has
$\Tr P_{I_L}(H_L) \ge 2$ for all sufficiently large $L$.

\ele

The main application is to the case where $\phi_i$ decay exponentially fast, so
Lemma \ref{lem:KM.Lemma.1} implies that for $L$ large enough, there are at least
two eigenvalues $\lam_1, \lam_2$ of the operator $H_L$ with
$\max \{|E -\lam_1|, |E -\lam_2| \} \le \half \eu^{-c L}$, for some $c>0$, hence with
$|\lam_1 - \lam_2| \le \eu^{-c L}$.

\proof[Proof of Theorem \ref{thm:simple.spectrum}]
Assume otherwise and fix any any $\th\in \ThinfM(g)$.
Since by construction
$\ThinfM(g) \subset\Thinf(g)$, $H(\om;\th)$ has pure point spectrum for every $\om\in\Om$,
and by Lemma \ref{lem:EF.unif.loc}, all its eigenfunctions decay exponentially fast.

Further, by construction of $\ThinfM(g)$, for any $\om\in\Om$ and all $j$ large enough,
all spectral spacings of $H_{\ball_{L_j}(0)}(\om;\th)$
are bounded from below by
$g\delta_j \ge L_j^{-\Const \ln L_j} = \eu^{-\Const \ln^2 L_j}$, and
$\eu^{-\Const \ln^2 L_j} > \eu^{-cL_j}$
for any $c>0$ and all $L_j$ large enough. This
contradicts Lemma \ref{lem:KM.Lemma.1} and proves the claim.
\qedhere

\vskip5mm

\par\noindent
\textbf{Acknowledgements.} It is a pleasure to thank Yakov Grigor'yevich Sinai,
Misha Goldstein and Abel Klein for numerous fruitful discussions of localization mechanisms
in deterministic disordered media; Tom Spencer for numerous discussions and warm hospitality
during my stay at the IAS in 2012; G\"{u}nter Stolz, Yulia Karpeshina and Roman Shterenberg
for stimulating discussions and  warm hospitality during my stay at the University
of Alabama at Birmingham in 2012.
%


\end{document}